\documentclass[useAMS,usenatbib]{mnras}

\usepackage[T1]{fontenc}
\usepackage{ae,aecompl}

\usepackage{graphicx}	
\usepackage{amsmath}	
\usepackage{amssymb}	
\usepackage{threeparttable}

\usepackage[latin1]{inputenc}  
\usepackage{color}
\usepackage{xspace}

\newcommand{\ox}{$\left[\ion{O}{\,\sc II}\right]$\xspace}
\newcommand{\oem}{$\left[\ion{O}{\,\sc II}\right]$~emitters\xspace}
\newcommand{\gl}{{\sc galform}\xspace}

\newcommand{\nm}{$\langle N \rangle_{M}$\xspace}
\newcommand{\nmo}{$\langle N \rangle_{\left[\ion{O}{ii}\right]}$\xspace}
\newcommand{\nmoc}{$\langle N \rangle_{\left[\ion{O}{ii}\right]\, cen}$\xspace}

\newcommand{\HII}{$\mathrm{H\textrm{\textsc{II}}}$\xspace}

\title[Model \oem at $0.5< z< 1.5$]{The host dark matter halos of \oem at $0.5< z< 1.5$}

\author[V. Gonzalez-Perez et al.]{
\parbox[c]{0.92\textwidth}{
\vspace{-0.8cm}
V. Gonzalez-Perez$^{1,2}$\thanks{E-mail: violegp@gmail.com (VGP)},
J. Comparat$^{3}$,
P. Norberg$^{2,4}$,
C. M. Baugh$^{2}$, 
S. Contreras$^{5}$, 
C. Lacey$^{2}$, 
N. McCullagh$^{2}$,
A. Orsi$^{6}$,
J. Helly$^{2}$, 
J. Humphries$^{2}$.}
\vspace*{6pt} 
\\
$^{1}$Institute of Cosmology \& Gravitation, University of Portsmouth, Dennis Sciama Building, Portsmouth, PO1 3FX, UK\\
$^{2}$Institute for Computational Cosmology, Department of Physics, Durham University, South Road, Durham, DH1 3LE, UK \\
$^3$ Max-Planck Institut fur extraterrestrische Physik, Postfach 1312, D-85741 Garching bei Munchen, Germany\\
$^{4}$ Centre for Extragalactic Astronomy, Department of Physics, Durham University, South Road, Durham, DH1 3LE, UK \\
$^{5}$ Instituto de Astrof\'isica, Pontifica Universidad Cat\'olica
de Chile, Santiago, Chile \\
$^{6}$ Centro de Estudios de F\'isica del Cosmos de Arag\'on, Plaza de San Juan 1, Teruel E-44001, Spain
}

\begin{document}
\label{firstpage}
\pagerange{\pageref{firstpage}--\pageref{lastpage}}
\maketitle

\date{Accepted XXX. Received YYY; in original form ZZZ}

\pubyear{2016}

\begin{abstract}
Emission line galaxies (ELGs) are used in several ongoing and upcoming surveys (SDSS-IV/eBOSS, DESI) as tracers of the dark matter distribution. Using a new galaxy formation model, we explore the characteristics of \oem, which dominate optical ELG selections at $z\simeq 1$. Model \oem at $0.5<z<1.5$ are selected to mimic the DEEP2, VVDS, eBOSS and DESI surveys. The luminosity functions of model \oem are in reasonable agreement with observations. The selected \oem are hosted by haloes with $M_{\rm halo}\geq 10^{10.3}h^{-1}{\rm M}_{\odot}$, with $\sim 90$\% of them being central star-forming galaxies. The predicted mean halo occupation distributions of \oem has a shape typical of that inferred for star-forming galaxies, with the contribution from central galaxies, \nmoc, being far from the canonical step function. The \nmoc can be described as the sum of an asymmetric Gaussian for disks and a step function for spheroids, which plateaus below unity. The model \oem have a clustering bias close to unity, which is below the expectations for eBOSS and DESI ELGs. At $z\sim 1$, a comparison with observed g-band selected galaxy, which are expected to be dominated by \oem, indicates that our model produces too few \oem that are satellite galaxies. This suggests the need to revise our modelling of hot gas stripping in satellite galaxies. 

\end{abstract}



\section{Introduction}
The quest to understand the nature of both dark matter and dark energy has led us to adopt new tracers of the large-scale structure of the Universe, such as emission line galaxies \citep[hereafter ELGs, e.g.\ ][]{dawson16,laureijs11,desi1, pozzetti16}. Current ELG samples are small and their characteristics are not well understood \citep{comparat16,kaasinen16}. Initial tests on relatively small area surveys indicate that there are enough ELGs to chart space-time and understand the transition between the dark matter and the dark energy dominated eras \citep{comparat13,okada16,delubac17}. Moreover, by measuring the properties of ELGs as tracers of star formation over a substantial amount of cosmic time, we can shed light on the mechanisms that quench the star formation in typical galaxies since the peak epoch of star formation around $z \simeq 2$ \citep{lilly96,madau96,mostek13}. 

The SDSS-IV/eBOSS\footnote{extended Baryon Oscillation Spectroscopic Survey,  \url{http://www.sdss.org/surveys/eboss/} \citep{dawson16}} survey is currently targeting what will become the largest sample to date of ELGs at $z\simeq 0.85$ \citep{comparat15eboss,raichoor17,delubac17}. This large sample will allow us to go beyond the current state-of-the-art cosmological constraints by measuring cosmological probes such as baryon acoustic oscillations and redshift space distortions at $z\sim 1$ \citep{zhao16}. This pioneering use of ELGs as cosmological probes is planned to be enhanced by future surveys, such as DESI\footnote{Dark Energy Spectroscopic Instrument, \url{http://desi.lbl.gov/} \citep{levi13}}, 
PFS\footnote{Prime Focus Spectrograph,\url{http://sumire.ipmu.jp/en/2652} \citep{takada14}}, 
WEAVE\footnote{Wide-field multi-object spectrograph for the William Herschel Telescope, \url{http://www.ing.iac.es/weave/} \citep{dalton14}}, 
and 4MOST\footnote{4-metre Multi-Object Spectroscopic Telescope, \url{https://www.4most.eu/} \citep{deJong14}}.

An ELG is the generic name given to any galaxy presenting strong emission lines associated with star-formation events. Galaxies with nuclear activity also present emission lines. However, the line ratios of such objects tend to be different from those driven by star formation activity because of the different ionisation states present \citep[e.g.][]{belfiore16}. The presence of these features allows for a robust determination of galaxy redshifts. Most of the sampled ELGs at $z\sim 1$ are expected to present a strong \ox line at a rest-frame wavelength of 3727 \AA. For detectors sampling optical to near infra-red wavelengths \oem can be detected up to $z=2$ \citep[e.g.][]{sobral12}. 

The fate of galaxies is determined by the growth of dark matter structures which, in turn, is affected by the nature of the dark energy. However, gravity is not the only force shaping the formation and evolution of galaxies. Baryons are affected by a multitude of other processes, mostly related to the fate of gas. Computational modelling is the only way we can attempt to understand all the processes involved in the formation and evolution of galaxies \citep[e.g.][]{somerville15}. The \ox emission is particularly difficult to predict since it depends critically on local properties, such as dust attenuation and the structure of the HII regions and their ionization fields. This is why \ox traces star formation and metallicity in a non-trivial way (e.g.\,\citealt{kewley04,mackay16}).

Previous work on modelling \oem has shown that semi-analytic galaxy formation models can reproduce their observed luminosity function (LF) at $z \sim 1$ \citep{orsi14,comparat15o2,comparat16}, making them ideal for studying the clustering properties of \oem and hence bias. These predictions are used in the design and interpretation of current and future surveys, such as eBOSS \citep{dawson16} and DESI \citep{desi1}. \citet{favole16} inferred the clustering and fraction of satellites for a g-band selected sample of galaxies that is expected to be dominated by ELGs at $0.6<z<1.7$. Their results are based on a modified sub-halo abundance matching (SHAM) technique that takes into account the incompleteness in the selection of ELGs, because not all haloes will contain an ELG. \citeauthor{favole16} found that their sample of g-selected galaxies at $z\sim 0.8$ is best matched by a model with $22.5\pm 2.5$\% of satellite galaxies and a mean host halo mass of $(1\pm 0.5) \times 10^{12}h^{-1} {\rm M_{\odot}}$. With the necessary modifications of the SHAM technique to provide a good description of the clustering of the observed ELGs, which is an incomplete sample of galaxies, the \nm for central ELGs is expected to differ from the canonical step function which reaches one central galaxy per halo, which is typical in mass limited samples.

Here we aim to characterise the nature of model \oem, as tracers of the star formation across cosmic time, and to study their expected mean halo occupation distribution and clustering to better understand \oem as tracers of the underlying cosmology. We adopt a physical approach rather than the empirical one used in \citet{favole16}. The use of a semi-analytical model of galaxy formation and evolution \citep[see][for some of the early developments in this field]{white91,lacey91,kauffmann93,cole94,som99} gives us the tools to understand the physical processes that are the most relevant for the evolution of ELGs in general and \oem in particular. Here we present a new flavour of \gl, developed based upon \citet{gp14}, a model that produced \oem LFs in reasonable agreement with observations \citep{comparat15o2}. 

The plan of this paper is as follows\footnote{The programs used to generate the plots presented in this paper can be foun in \url{https://github.com/viogp/plots4papers/}}. In \S~\ref{sec:sams} we introduce a new galaxy model (GP17), which is an evolution of previous \gl versions. In \S~\ref{sec:lf} the \ox luminosity functions from different observational surveys are compared to model \oem selected to mimic these surveys. These selections are explored in both \S~\ref{sec:exploring} and Appendix~\ref{sec:colours}. Given the reasonable agreement found between this GP17 model and current observations, we infer the mean halo occupation distribution in \S~\ref{sec:hod}, and clustering in \S~\ref{sec:xi} of \oem. In \S~\ref{sec:conclusions} we summarise and discuss our results.

\section{The Semi-Analytic model}\label{sec:sams}



Semi-analytical (SA) models use simple, physically motivated rules to follow the fate of baryons in a universe in which structure grows hierarchically through gravitational instability \citep[see][for an overview of hierarchical galaxy formation models]{baugh06,benson10}. 

\gl was introduced by \citet{cole00} and since then it has been enhanced and improved \citep[e.g.][]{baugh05,bower06,fanidakis11,lagos11,lacey16}. \gl follows the physical processes that shape the formation and evolution of galaxies, including: (i) the collapse and merging of dark matter haloes; (ii) the shock-heating and radiative cooling of gas inside dark matter haloes, leading to the formation of galaxy discs; (iii) quiescent star formation in galaxy discs which takes into account both the atomic and molecular components of the gas \citep{lagos11}; (iv) feedback from supernovae, from active galactic nuclei \citep{bower06} and from photo-ionization of the intergalactic medium; (v) chemical enrichment of the stars and gas (assuming instantaneous recycling); (vi) galaxy mergers driven by dynamical friction within common dark matter haloes. \gl provides a prediction for the number and properties of galaxies that reside within dark matter haloes of different masses. 

Currently there are two main branches of \gl: one with a single initial mass function \citep[IMF][hereafter GP14]{gp14} and one that assumes different IMFs for quiescent and bursty episodes of star formation \citep{lacey16}. 

Here we introduce a new version of the \gl model of the formation and evolution of galaxies (hereafter GP17), which will be available in the Millennium Archive Database\footnote{\url{http://www.virgo.dur.ac.uk/}, \\ \url{http://gavo.mpa-garching.mpg.de/Millennium/}}. The details specific to the GP17 model are introduced in \S~\ref{sec:gp17}. Below we also give further details of how \gl models emission lines, \S~\ref{sec:sams_lines},  and dust, \S~\ref{sec:sams_dust}, as these are key aspects to understand the results from this study.

\subsection{The GP17 model}\label{sec:gp17}
\begin{table}
\caption{
Differences between the GP17 and the GP14 \gl implementation. 
$\alpha_{\rm cool}$ is one of the parameters setting the AGN feedback efficiency in \gl \citep[Eq.~12 in][]{lacey16} and $v_{\rm SN}$ is related to the modelling of SN feedback \citep[Eq.~10 in][]{lacey16}.}
\hspace*{-0.7cm}
\begin{threeparttable}[t]
\label{tbl:galform}
\begin{center}
\begin{tabular}{|c|c|c|}
\hline
\gl parameter & GP14 & GP17 (this work) \\
\hline
   IMF\tnote{a} & \citeauthor{kennicutt_imf} & \citeauthor{cha03} \\
   SPS model   &  BC99\tnote{b} & \citeauthor{cw09}\\
   Stripping of hot gas   &   instantaneous & gradual \\
   Merging scheme   &   \citeauthor{lacey93} & \citeauthor{simha16} \\
   $\alpha_{\rm cool}$ (AGN feedback) &    0.6 & 0.9 \\
   $v_{\rm SN}$ (SN feedback) [km/s] &    425 & 370 \\
\hline
\end{tabular}
\begin{tablenotes}
\item[a] The metal yield and recycled fractions are not considered as free parameters here since their values are set by the assumed stellar IMF, following calculations carried out with {\sc pegase2} \citep{pegase2}. In GP17 we fix the metal yield to 0.02908 and the recycled fraction to 0.4588.\\
\item[b] BCC99 is an updated version of the \citet{bc93} SPS model.
\end{tablenotes}
\end{center}
\end{threeparttable}
\end{table}

The GP17 model uses dark matter halo merger trees extracted from the MS-W7 N-body simulation \citep{guo13,jiang14,gp14}, a box of 500 $h^{-1}$Mpc aside and with a cosmology consistent with the 7$^{\rm th}$ year release from WMAP \citep{wmap7}: matter density $\Omega_{m, 0}=0.272$, cosmological constant $\Omega_{\Lambda , 0}=0.728$, baryon density $\Omega_{b,0}=0.0455$, a normalization of density fluctuations given by $\sigma_{8, 0}=0.810$ and a Hubble constant today of $H(z=0) = 100 h{\rm km\, s}^{-1}{\rm Mpc}^{-1}$ with $h=0.704$. 

The model in this study, GP17, assumes a single IMF, building upon the \gl versions presented in both GP14 and \citet{g16}. The main two aspects that are different in the GP17 model with respect to GP14 are: i) the assumption of a gradual stripping of the hot gas when a galaxy becomes a satellite by merging into a larger halo \citep{font08,lagos14.h1} and ii) the use of a new merging scheme to follow the orbits of these satellite galaxies \citep{simha16}. 

Table~\ref{tbl:galform} summarizes all the differences between the new GP17 model and the GP14 \citep{gp14} \gl implementation. We review below the changes made in the same order as they appear in Table~\ref{tbl:galform}.

The GP17 model assumes the IMF from \citet{cha03}. This IMF is widely used in observational derivations, and thus this choice facilitates a more direct comparison between the model results and observational ones. GP17 uses the flexible \citet{cw09} stellar population synthesis (SPS) model (CW09 hereafter). Coupling this SPS model to \gl gives very similar global properties for galaxies over a wide range of redshifts and wavelengths to using the \citet{bc03} SPS model \citep[as in][]{gp14}. The CW09 SPS model was chosen here over that of \citet{bc03} because it provides greater flexibility to explore variations in the stellar evolution assumptions.

\subsubsection{The treatment of gas in satellite galaxies}
\begin{figure}  
\includegraphics[width=0.5\textwidth]{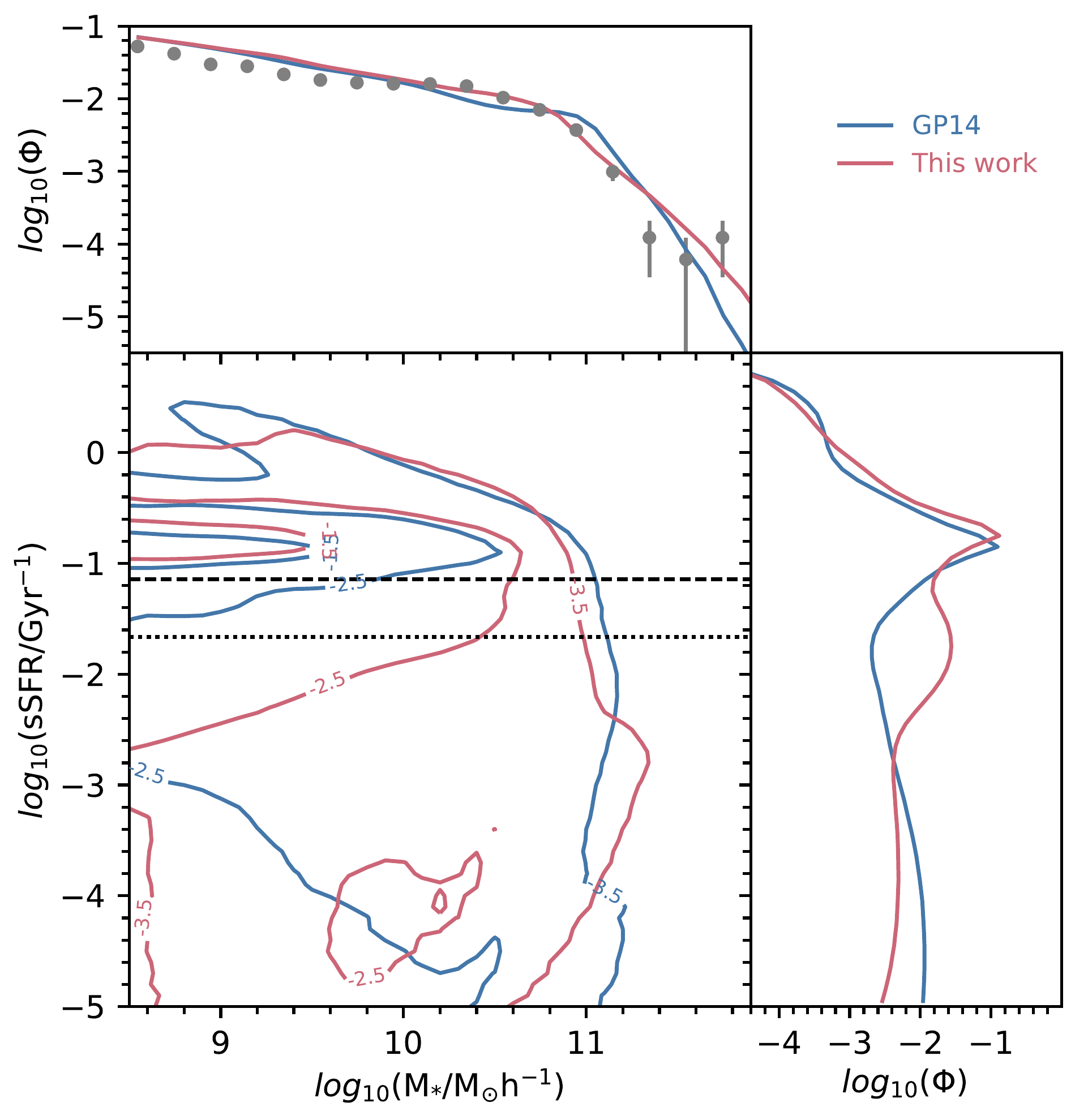}
\caption{\label{fig:gsmf} The $z=0$ distribution of galaxies in the sSFR-stellar mass plane for galaxies from the model presented here (red lines) and that described in \citet{gp14} (blue lines). The horizontal dotted line corresponds to the boundary proposed by \citet{franx08} to separate star forming from passively evolving galaxies, $ {\rm sSFR} = 0.3/t_{\rm Hubble}(z)$, while the dashed line simply shows ${\rm sSFR}= 1/t_{\rm Hubble}(z)$, for comparison. The sSFR-stellar mass plane has been collapsed into the galaxy stellar mass function, top, and the sSFR function, right. The corresponding densities shown are $\Phi (h^{3}{\rm Mpc}^{-3}{\rm dex}^{-1})$. The $z=0$ stellar mass function is compared to results from \citet{baldry12}, grey symbols. Following \citet{lacey16}, the estimations from the GP14 model have been corrected from the assumed Kennicutt IMF to the \citeauthor{cha03} one assumed in both observations and the GP17 model. 
} 
\end{figure}
\begin{figure} 
\includegraphics[width=0.45\textwidth]{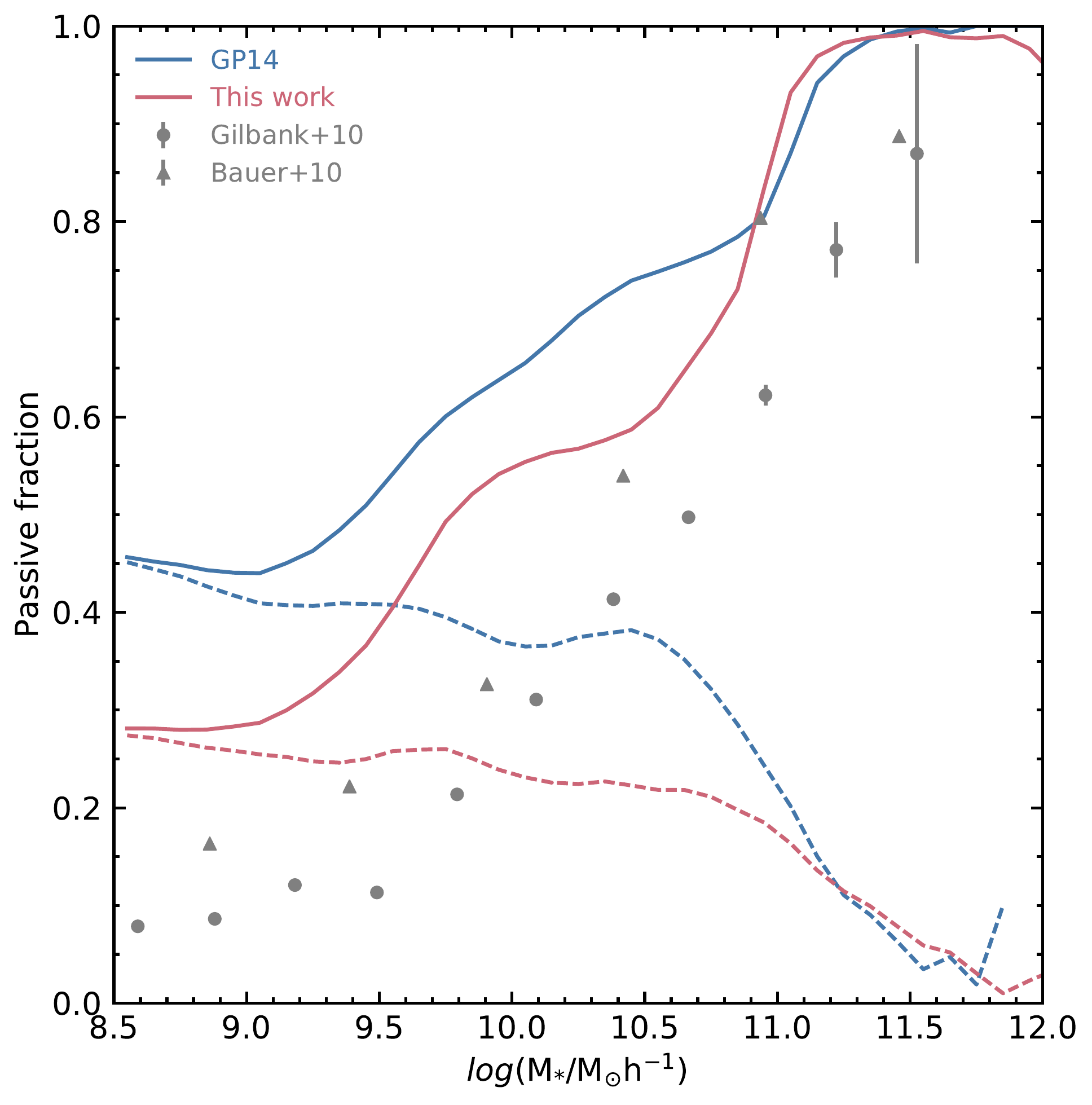}
\caption{\label{fig:pf_cal} The fraction of passive galaxies at $z=0$, i.e.\ those with sSFR $<0.3/t_{\rm Hubble}(z=0)$, in the GP17 model (solid red line) and in \citet{gp14} (solid blue line), compared to the observational results from \citet{gilbank10} and  \citet{bauer13} (triangles), as extracted and presented in \citet{furlong15}. The dashed lines show the contribution of satellite galaxies to the total passive fraction.
} 
\end{figure}

In the GP17 model the hot gas in satellites is removed gradually, using the model introduced by \citet{font08} based on a comparison to hydrodynamical simulations of cluster environments \citep{mccarthy08}. This change has a direct impact on the distribution of specific star formation rates. Compared to the GP14 model, some galaxies from the GP17 model have higher sSFR values, in better agreement with observational inferences \citep{weinmann09}. This is clearly seen in the sSFR function around $\log_{10} \left( {\rm sSFR} / {\rm Gyr}^{-1} \right) \sim -1.5$ presented in Fig.~\ref{fig:gsmf}. This choice reduces the fraction of passive model galaxies with $M_{*}< 10^{11} h^{-1} {\rm M_{\odot}}$. As shown in Fig.~\ref{fig:pf_cal},  the resulting a passive fraction is closer to the observational results at $z=0$, compared with models such GP14, which assumes instantaneous stripping of the hot gas from satellites \citep[see also][and discussions therein]{lagos14,g16}. Note that we have not made a direct attempt to reproduce the observed passive fraction by adjusting the time scale for the hot gas stripping in satellite galaxies, but rather we have simply used the parameters introduced in \citet{font08}. We leave a detailed exploration of the effect of environmental processes on galaxy properties for another study. The passive fraction at $z=0$ is obtained using the limit on the specific star formation rate, $ {\rm sSFR} = {\rm SFR}/M_{*}$, proposed in \citet{franx08}, i.e.\, $ {\rm sSFR} < 0.3/t_{\rm Hubble}(z)$, where $t_{\rm Hubble}(z)$ is the Hubble time, $t_{\rm Hubble}=1/H$, at redshift $z$. Fig.~\ref{fig:gsmf} shows the $z=0$ distribution of the sSFR and stellar mass for GP17 model galaxies, together with those from the GP14 model, compared to the limits $ {\rm sSFR} = 0.3/t_{\rm Hubble}(z)$ and $ {\rm sSFR } < 1/t_{\rm Hubble}(z)$ (horizontal dotted and dashed lines respectively). The contours show that the main sequence of star-forming galaxies, i.e.\, the most densely populated region in the sSFR-$M_{*}$ plane, is above both these limits, while passively evolving galaxies, i.e.\ those with low star formation rates, are below them. Fig.~\ref{fig:gsmf} also shows the model galaxy stellar mass function at $z=0$ compared with observations.

\subsubsection{The merging scheme for satellite galaxies}
The GP17 model is the first publicly available \gl model to use the new merging scheme introduced by \citet{simha16}. In this merging scheme, satellite galaxies associated with resolved sub-haloes cannot merge with the central galaxy until their host sub-halo does. Satellite galaxies with no associated resolved sub-halo merge with their central galaxy after a time calculated analytically, taking into account dynamical friction and tidal stripping. As described in \citet{campbell15}, compared to observations up to $z=0.7$, the radial distribution of \gl galaxies is too highly concentrated \citep[see also][]{contreras13}. As a result of using the merging scheme of \citeauthor{simha16}, satellite galaxies merge more quickly  with their central galaxy than it was previously assumed by the analytical function used \citep{lacey93}. This, along with the  modification to the radial distribution of satellite galaxies results in an improved match to the observed two point correlation function at small scales \citep{campbell15}. 

\subsubsection{Calibration of the free parameters}
\begin{figure}
\includegraphics[width=0.5\textwidth]{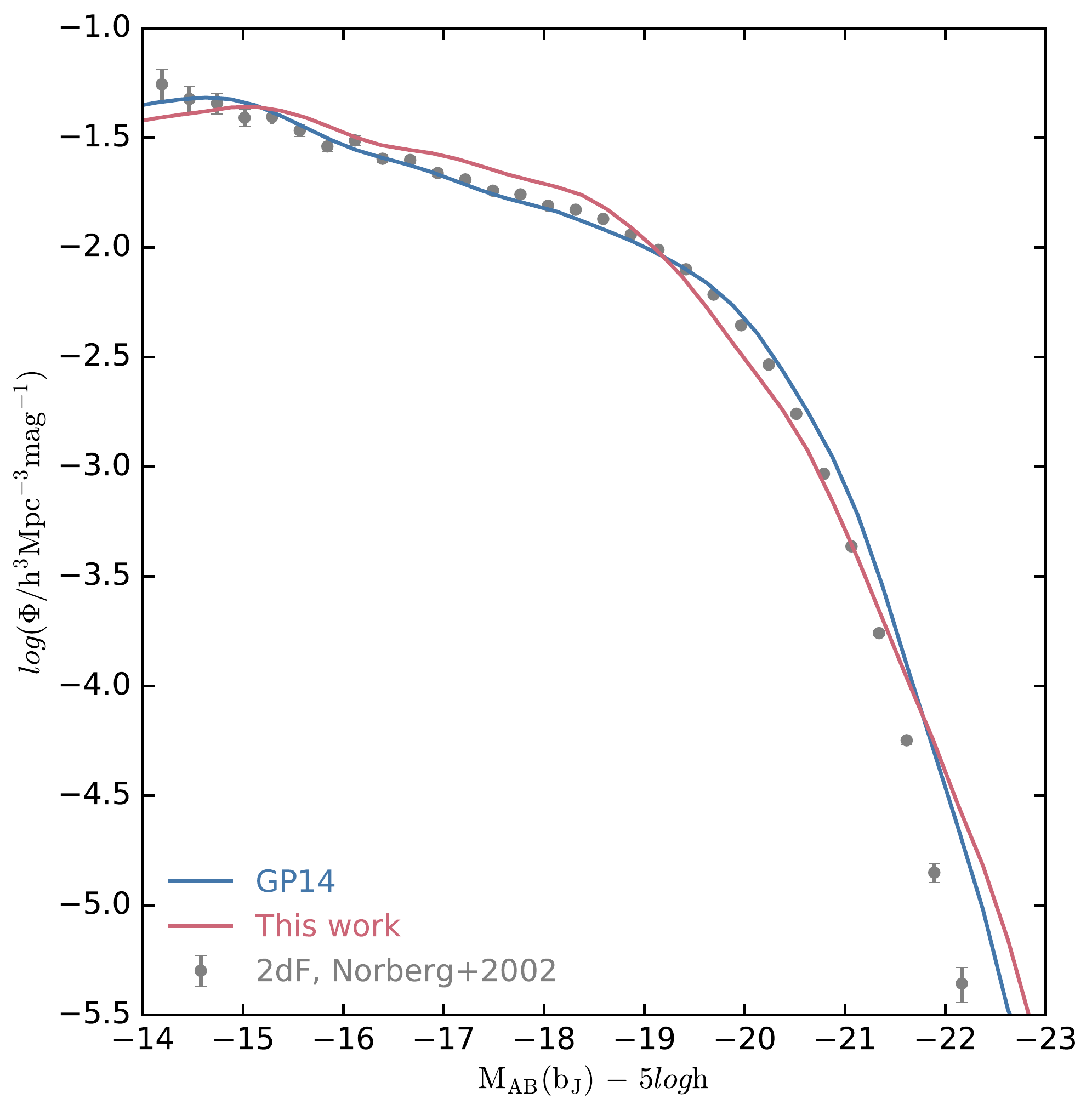}
\includegraphics[width=0.5\textwidth]{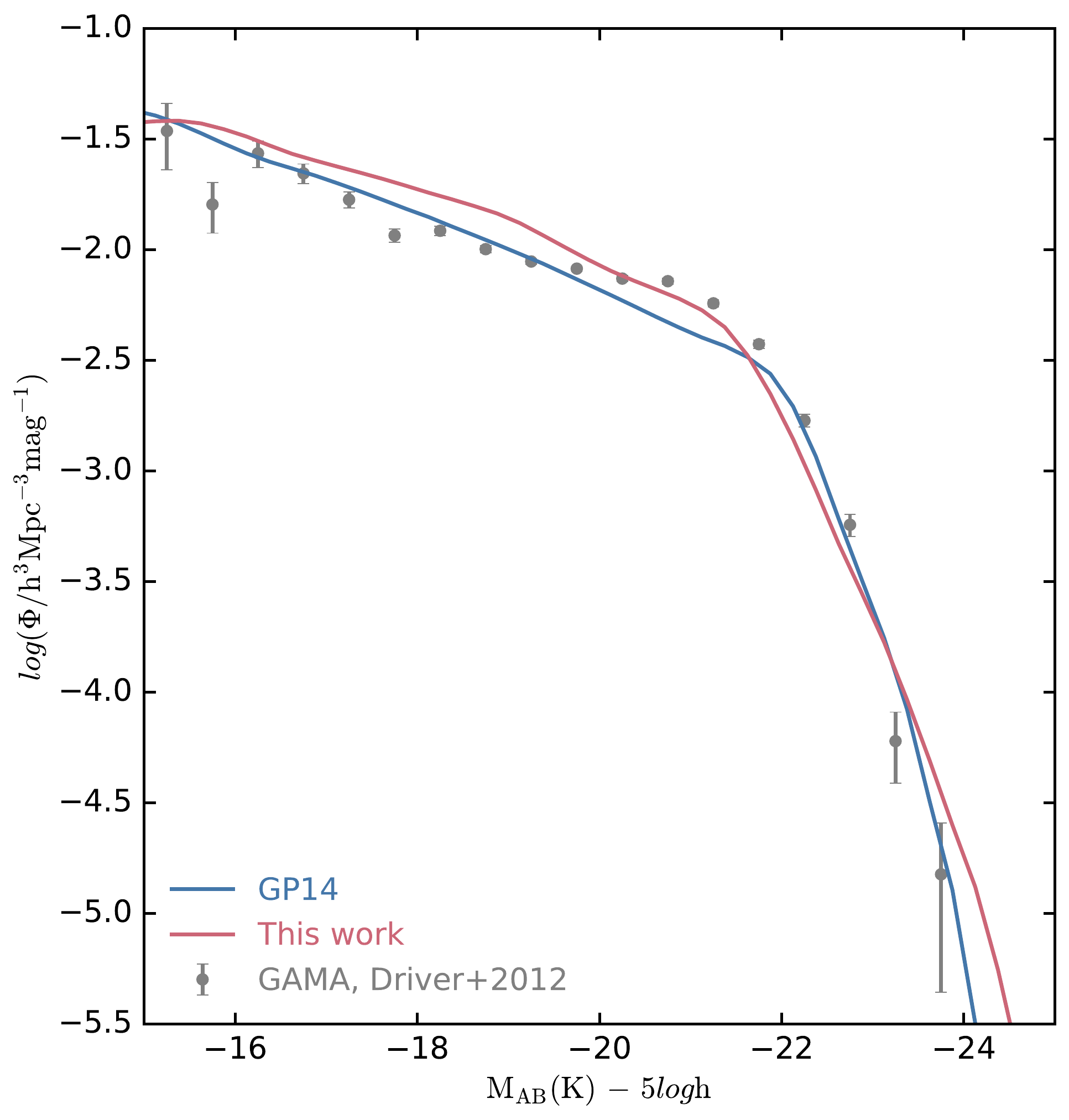}
\caption{\label{fig:lfcal} The predicted luminosity functions at $z=$0 (solid lines), in the b$_{\rm J}$-band ($\lambda _{\rm eff}= 4500$\AA, top) and in the K-band ($\lambda _{\rm eff}=2.2\mu$m, bottom), compared with observations from \citet{norberg02} and \citet{driver12}, respectively. The blue lines show the predictions from the GP14 model, while the red lines show the predictions from the new GP17 model presented here. These data were used to calibrate the free parameters of the model.
 } 
\end{figure}
The free parameters in the GP17 model have been calibrated to reproduce the observed luminosity functions\footnote{Note that throughout this work all quoted magnitudes are in the AB system, unless specified otherwise.} at $z=0$ in both the b$_J$ and K-bands \citep{norberg02,driver12}, as shown in Fig.~\ref{fig:lfcal}, to give reasonable evolution of the UV and V-band luminosity functions and to reproduce the observed black hole-bulge mass relation (not shown here but which matches observations equally well as in GP14). When calibrating the GP17 model, our aim was to make the smallest number of changes to the GP14 model parameters. A side effect of incorporating the merging scheme from \citeauthor{simha16} into the model is an increase of the number of massive central galaxies at $z=0$, that has to be compensated for by modifying the galactic feedback, in order to recover the same level of agreement with the observational datasets used during the calibration of the model  parameters. To achieve this, both the efficiency of the supernova feedback and the mass of haloes within which gas cooling stops due to AGN feedback have been reduced. The changes to these two parameters related to galactic feedback allow for the GP17 model to match the observed $z=0$ luminosity functions shown in Fig.~\ref{fig:lfcal} with a $\chi ^2$ that is just a factor of 3 larger than that for luminosity functions from the GP14 model.

\subsection{The emission line model}\label{sec:sams_lines}

The GP14 model predicts the evolution of the H${\alpha}$ LF reasonably well \citep{lagos14}. H${\alpha}$ is a recombination line and thus its unattenuated luminosity is directly proportional to the number of Lyman continuum photons, which is a direct prediction of the \gl model \citep{orsi08,orsi10}. The main uncertainty in the case of the H${\alpha}$ line is the dust attenuation.

In \gl, the ratio between the \ox luminosity and the number of Lyman continuum photons is calculated using the \HII region models of \citet{sta90}. The \gl model uses by default eight \HII region models spanning a range of metallicities but with the same uniform density of 10 hydrogen particles per cm$^{-3}$ and one ionising star in the center of the region with an effective temperature of 45000 K. The ionising parameter\footnote{Following \citet{sta90}, the ionising parameter at a given radius, $r$, of the HII region, $U(r)$ is defined here as a dimensionless quantity equal to the ionizing photon flux, $Q$, per unit area per atomic hydrogen density, $n_{\rm H}$, normalised by the speed of light, c: $U(r)=Q/(4\pi r^2 n_{\rm H} c)$.} of these \HII region models is around $10^{-3}$, with exact values depending on their metallicity in a non-trivial way. These ionising parameters are typical within the  grid of \HII regions provided by \citet{sta90}. 

In this way, the \gl model assumes a nearly invariant ionization parameter. This assumption, although reasonable for recombination lines, is possibly too simplistic in practice for other emission lines such as \ox \citep[e.g.][]{sanchez14}. Nevertheless, with this caveat in mind, we shall study the predictions of \gl for \oem with the simple model here and defer the use of a more sophisticated emission line model to a future paper. 

We have also run the galaxy formation model together with the empirical emission line ratios described in \citet{anders03}. These authors provide line ratios for 5 metallicities, combining the observational database of \citet{izotov94,izotov97} and \citet{izotov98} for $Z=0.0004$ and $Z=0.004$, and using \citet{sta90} models for higher metallicities, $Z=0.008,\, 0.02,\, 0.05$. \citet{anders03} provide line ratios with respect to the flux of the H$_{\beta}$ line, which they assume to be $4.757 \times 10^{-13}$ times the number of hydrogen ionising photons, $N_{\rm Lyc}$. For the other Hydrogen lines we assume the low-density limit recombination Case B (the typical case for nebulae with observable amounts of gas) and a temperature of 10000 K: $F(Ly_{\alpha})/F(H_{\beta})=32.7$, $F(H_{\alpha})/F(H_{\beta})=2.87$, $F(H_{\gamma})/F(H_{\beta})=0.466$ \citep[Tables 4.1 and 4.2 in][]{osterbrock06}. These line ratios have been reduced by a factor of 0.7 for gas metallicities $Z \ge 0.08$ to account for absorption of ionising photons within the HII region \citep{anders03}. 

We have done all the analysis presented in this paper using the \citet{anders03} model for HII regions obtaining very similar results to those presented below when using the default models from \citet{sta90}. Thus, all the conclusions from this work are also adequate when the \citet{anders03} models are assumed.

\subsection{The dust model}\label{sec:sams_dust}
Emission lines can only be detected in galaxies that are not heavily obscured and thus, survey selections targetting ELGs are likely to miss dusty galaxies. In \gl the dust is assumed to be present in galaxies in two components: diffuse dust (75\%) and molecular clouds (25\%). This split is consistent, within a factor of two, with estimates based on observations of nearby galaxies \citep{granato00}. The diffuse component is assumed to follow the distribution of stars. Model stars escape from their birth molecular clouds after 1 Myr (the metallicity is assumed to be the same for the stars and their birth molecular clouds). Given the inclination  of the galaxy and the cold gas mass and metallicity, the attenuation by dust at a given wavelength is computed using the results of a radiative transfer model \citep[see][for further details on the modelling of dust attenuation]{gp13}.

Lines are assumed to be attenuated by dust in a similar way to the stellar continuum, as described above. Thus, the predicted  \ox luminosity should be considered as an upper limit as some observational studies find that the nebular emission of star forming galaxies experiences greater (by up to a factor of 2) dust extinction than the stellar component \citep{calzetti97,debarros16}. Nevertheless, given the uncertainty in the dust attenuation at the redshifts of interest, the line luminosities are calculated using the model stellar continuum dust attenuation. It is worth noting that, as was also found for cluster galaxies at $z\sim 1$ \citep{merson16}, less  than 3\% of the model ELGs (mostly the brightest \oem) are attenuated by more than one magnitude in the rest frame NUV to optical region of the spectra, which is due to the very small sizes and large cold gas content of those galaxies.

\section{Model $[$OII$]$ emitters}
\begin{table*}
  \caption{The cuts applied to the model galaxies in order to mimic the selection of \oem in the corresponding observational survey, following the results from \citet{comparat15o2}. For \oem, DEEP2 covers the redshift range of $0.7<z<1.3$ and VVDS spans $0.5<z<1.3$ \citep{comparat15o2}. Low redshift galaxies are avoided in the DEEP2 survey by imposing a colour-colour cut. However, here we simply make a cut in the studied redshift. For the case of the eBOSS and DESI selections, very blue \oem at the target redshift range are discarded due to stellar contamination (further details can be found in Appendix~\ref{sec:colours}). Thus, we apply here the colour cuts described in \citet{comparat15eboss} for the eBOSS selection 
and those described in \citet{desi1} for the DESI selection. The magnitudes are on the AB system. The particular filter response used for the different cuts is indicated by a superscript on the magnitude column. 
}
\begin{center}
\begin{tabular}{|c|c|c|c|}
\hline
Cuts to & Apparent & \ox flux & Colour\\
mimic  & magnitude & (${\rm erg\, s^{-1}cm^{-2}}$) & selection\\
\hline \vspace{0.1cm}

DEEP2  & $R_{AB}^{\rm DEIMOS}<24.1$ & $2.7\times 10^{-17}$ & None\\ \vspace{0.1cm}

VVDS-Deep &  $i_{AB}^{\rm CFHT} \leq 24$ & $1.9\times 10^{-17}$ & None \\ \vspace{0.1cm}

VVDS-Wide &  $i_{AB}^{\rm CFHT} \leq 22.5$ & $3.5\times 10^{-17}$ & None \\ \vspace{0.1cm}

eBOSS  &  $22.1< g_{AB}^{\rm DECam} < 22.8$ & $1\times 10^{-16}$ & $0.3< (g-r) < 0.7$  \& \\ 
     & & & $0.25 < (r-z) < 1.4$  \&  \\ \vspace {0.2cm}
     & & & $0.5(g-r)+0.4 < (r-z) < 0.5(g-r)+0.8$ \\  

DESI &  $r_{AB}^{\rm DECam} < 23.4$  & $8\times 10^{-17}$ & $(r-z) > 0.3$  \& $(g-r) > -0.3$ \& \\
     & & & $0.9(g-r)+0.12 < (r-z) < 1.345 - 0.85(g-r)$ \\
\hline
\end{tabular}
\end{center}
\label{tbl:obs}
\end{table*}

Star forming galaxies exhibiting strong spectral emission lines are generically referred to as emission line galaxies (ELGs). Present and future surveys such as eBOSS, Euclid and DESI target galaxies within a particular redshift range \citep{laureijs11,dawson16,desi1,desi2}. The specific redshift range and the type of detectors used by a survey will determine which spectral lines will be observed. We focus here on those surveys with optical and near-infrarred detectors targeting ELGs at $z\sim 1$ which will have prominent \ox lines. We will refer to these galaxies as \oem. 

%
Within the redshift range $0.6 \le z\le 1.5$, at most 10\% of all model galaxies are \oem, following the definitions of Table~\ref{tbl:obs} (see \S~\ref{sec:selection}). This percentage depends on the minimum galaxy mass, which in this case is set by the resolution of the simulation used. Over 99\% of model \oem are actively forming stars (as defined in \S~\ref{sec:sams}) and over 90\% are central galaxies. 

In \S~\ref{sec:selection} we describe how we select model \oem. We compare the model luminosity functions at $0.6 \le z\le 1.5$ with observations in \S~\ref{sec:lf} and we explore the selection properties in \S~\ref{sec:exploring}.

\subsection{The selection of $[$OII$]$ emitters}\label{sec:selection}
\oem are selected from the model output to mimic the set of surveys summarised in the first column of Table~\ref{tbl:obs}. The DEEP2 survey used the Keck DEIMOS spectrograph to obtain spectra of $\sim$50,000 galaxies in four separate fields covering $\sim$2.8 deg$^2$ \citep{newman13}. The VVDS survey was conducted using the VIMOS multi-slit spectrograph on the ESO-VLT, observing galaxies up to $z=6.7$ over 0.6 deg$^2$ for the Deep survey and 8.6 deg$^2$ in the Wide one \citep[VIMOS VLT Deep Survey Database][]{lefevre13}.  The eBOSS survey is on-going while the DESI survey has not yet started. Appendix~\ref{sec:colours} details the eBOSS and DESI selections, summarised in Table~\ref{tbl:obs}. Model \oem are selected using the same observational magnitude cuts (second column in Table~\ref{tbl:obs}), omitting colour cuts designed to remove low redshift galaxies as in this study galaxies are selected from the relevant redshift range already. For eBOSS and DESI selections we include the colour cuts designed to remove stellar contaminants (fourth column in Table~\ref{tbl:obs}; see Appendix~\ref{sec:colours} for further details). A limit on \ox flux has been added (third column in Table~\ref{tbl:obs}) to select model galaxies with a completeness that mimics the constraints from observational surveys. 

In the redshift range considered, over 85\% of all model galaxies are found to be star-forming. From these, a very small percentage, less than $1$\% in most cases, is classified as \oem by the restrictive VVDS-Wide, eBOSS and DESI selections. Given the mass resolution of our model, for the VVDS-Deep and DEEP2 cuts, \oem account for at most 11\% of the total star forming population at $z=0.62$, and this percentage decreases with increasing redshift. 

\subsection{Luminosity functions}\label{sec:lf}
\begin{figure}  \includegraphics[width=0.5\textwidth]{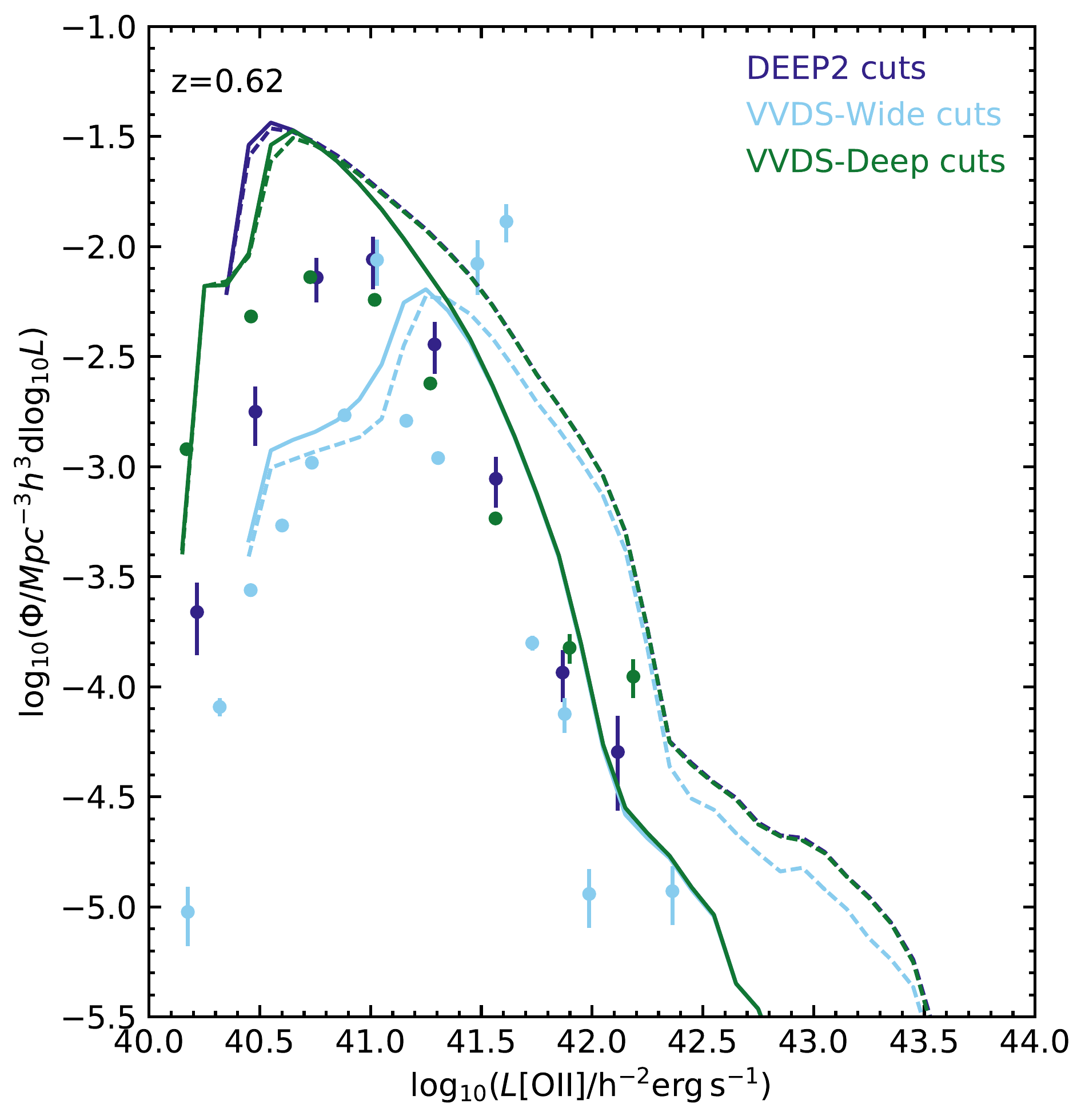}
\caption{\label{fig:lf} The LF of \oem at $z=0.62$ for model galaxies selected with the DEEP2 (dark blue, mostly over-plotted), VVDS-Wide (light blue) and VVDS-Deep (green) cuts given in Table~\ref{tbl:obs}. The solid lines present the model \ox dust attenuated luminosity function, while the dashed lines show the intrinsic luminosity function without considering dust attenuation. Observational data from \citet{comparat16} are shown as filled circles, with colours matching the cuts used to mimic the corresponding survey selection, as indicated in the legend. The observational errors come from jackknife re-sampling \citep{comparat16} and in some cases are smaller than the symbol.} 
\end{figure}
\begin{figure}
\includegraphics[width=0.5\textwidth]{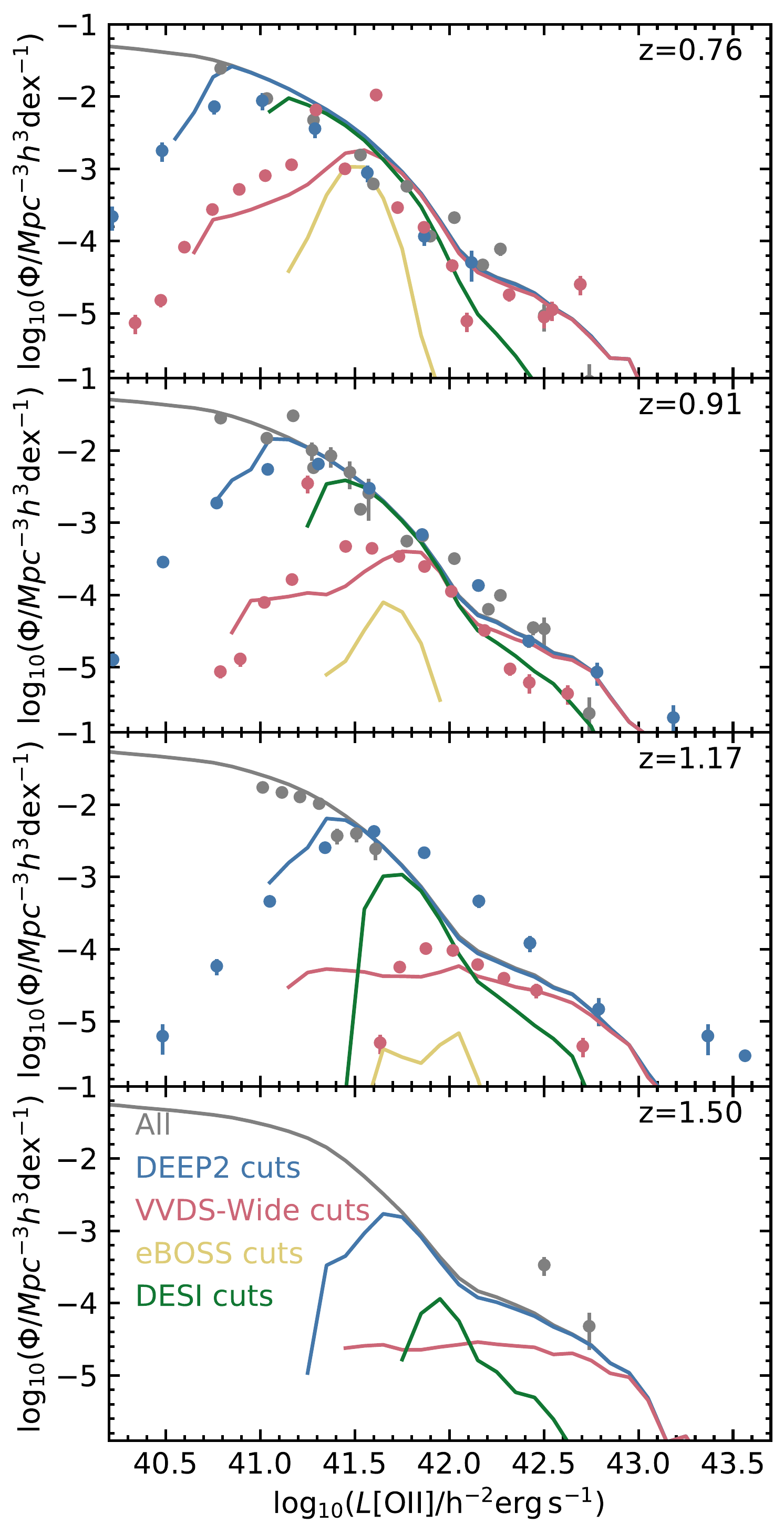}
\caption{\label{fig:lfall} From top to bottom: the LF of \oem at $z=0.76$, $0.91$, $1.17$ and $1.50$, for all model galaxies (grey lines), and those selected imposing the Table~\ref{tbl:obs} cuts mimicking DEEP2 (blue), VVDS-Wide (red), eBOSS (yellow), and DESI (green). The grey circles show the observed \oem LF constructed using complete data in a particular luminosity bins by \citet{comparat16}. Data from DEEP2 and VVDS are colour coded like the model galaxies selected to mimic both surveys. The observational errors come from jackknife re-sampling \citep{comparat16} and in some cases are smaller than the symbol.} 
\end{figure}
The luminosity function for \oem at $z=0.62$ from the GP17 model is compared in Fig.~\ref{fig:lf} to the observational compilation done by \citet{comparat16}\footnote{\url{http://projects.ift.uam-csic.es/skies-universes/SUwebsite/index.html}}, that includes data from the VVDS-Deep, VVDS-Wide and DEEP2 surveys among others \citep{ly07,gilbank10,sobral12,drake13,ciardullo13,khostovan15}. The model LFs are in reasonable agreement with the observations, with differences within a factor of 5 for densities above $10^{-5}{\rm Mpc}^{-3}h^3dex^{-1}$. Given the similarities between the GP17 and GP14 models, this was expected, as the GP14 model was already shown to be in reasonable agreement with observations at $z\sim1$ \citep{comparat15o2}. 

Galaxies with an ongoing star-burst dominate the bright end of the model \ox LF, $L {\rm [OII]}>10^{42} h^{-2} {\rm erg\, s^{-1}}$, producing the change in the slope of the LF seen at low number densities in Fig.~\ref{fig:lf}. The bright end of the LFs of \oem selected with the DEEP2, VVDS-Wide and VVDS-Deep cuts are also dominated (at the $\sim$80\% level) by spheroid galaxies (i.e.\ those with a bulge to total mass above 0.5). More than half of these \oem with $L{\rm[OII]}>10^{42} h^{-2} {\rm erg\, s^{-1}}$ have half mass radii smaller than $3 h^{-1} {\rm kpc}$. 

The DEEP2 cut uses the DEIMOS R-band filter response, while the VVDS cuts use the CFHT MegaCam\footnote{\url{http://svo2.cab.inta-csic.es/svo/theory/fps3/index.php?mode=browse\&gname=CFHT\&gname2=MegaCam}} i-band filter response. The value of the luminosity at which there is a turnover in the number of faint galaxies is sensitive to the particular filter response used. Above this luminosity, the model LF changes by less than 0.1 dex in number density if the R or i bands from DEIMOS, CFHT, PAN-STARRS or DES camera are used (note that not all this bands are used for Fig.~\ref{fig:lf}.

We note that BC99, an updated version of the \citet{bc93} SPS model, is used by default in most \gl models. As the spectral energy distribution below 912\AA~can widely vary among different SPS models \citep{gp14}, we verified that using the CW09 SPS model (as done in GP17) has a negligible impact on the luminosity function of \oem . Finally, Fig.~\ref{fig:lf} shows that at $z=0.62$ the model reproduces reasonably well the observed LF for \oem, including the decline in numbers due to the corresponding flux limits (summarised in Table~\ref{tbl:obs}). 

In Fig.~\ref{fig:lfall} the predicted LFs of \oem are shown at $z=0.76$, $0.91$, $1.17$ and $1.5$, sampling the relevant redshift ranges of the current eBOSS and future DESI surveys. The \oem LFs include effects from the five selection criteria summarised in Table~\ref{tbl:obs} and are compared to the observational data compiled by \citet{comparat16}. Fig.~\ref{fig:lfall} shows only the LFs with dust attenuation included. As noted before, since the model assumes the same attenuation for the emission lines as for the stellar continuum, the predicted \ox luminosity functions in  Fig.~\ref{fig:lfall} should be considered as overestimates \citep{calzetti97,debarros16}. Nevertheless, the predicted continuum extinction is significant and arguably larger than suggested by observations at $z\sim 1$, which might compensate for the lack of any additional attenuation being applied to the emission line 
luminosities in the model.


\subsection{Exploring the ELG selection}\label{sec:exploring}
\begin{figure}  
\includegraphics[width=0.5\textwidth]{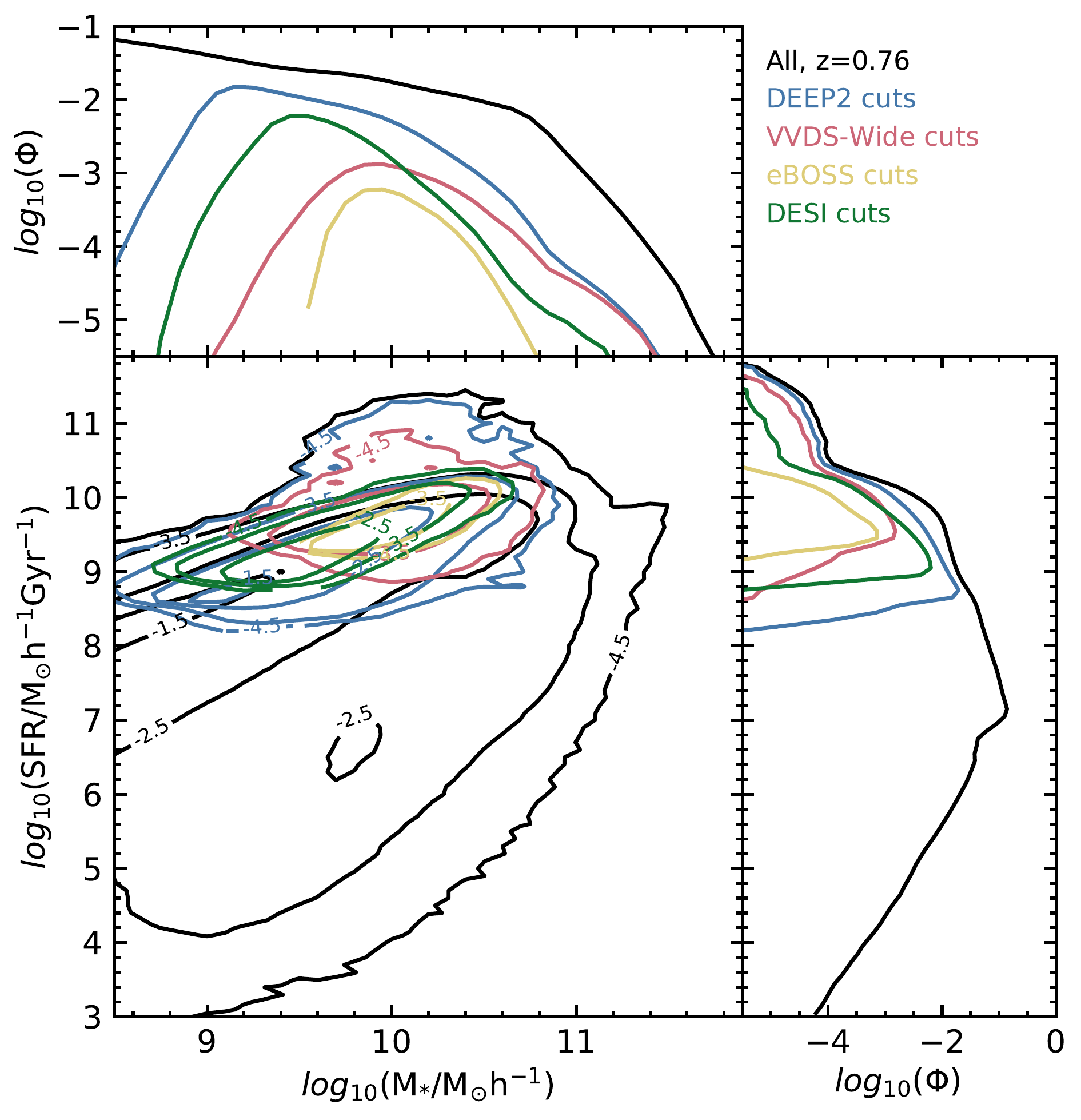}
\caption{\label{fig:gsmf_all} The galaxy distribution in the star-formation rate (SFR) - stellar mass (M$_{\star}$) plane at $z=0.76$: the density contours, shown as log$_{10} (\Psi / h^{3} {\rm Mpc}^{-3} {\rm dex}^{-2})$, are for all galaxies (black) and for survey specific selections, each colour coded following the key. The SFR-M$_{\star}$ plane has been collapsed into the galaxy stellar mass function, top panel, and the SFR function, right panel, with $\Phi$ in units of $h^{3} {\rm Mpc}^{-3} {\rm dex}^{-1}$. Basically all surveys selections result primarily in the inclusion of only star-forming model galaxies, but with different levels of sample completeness.} 
\end{figure}
\begin{figure}  
\includegraphics[width=0.5\textwidth]{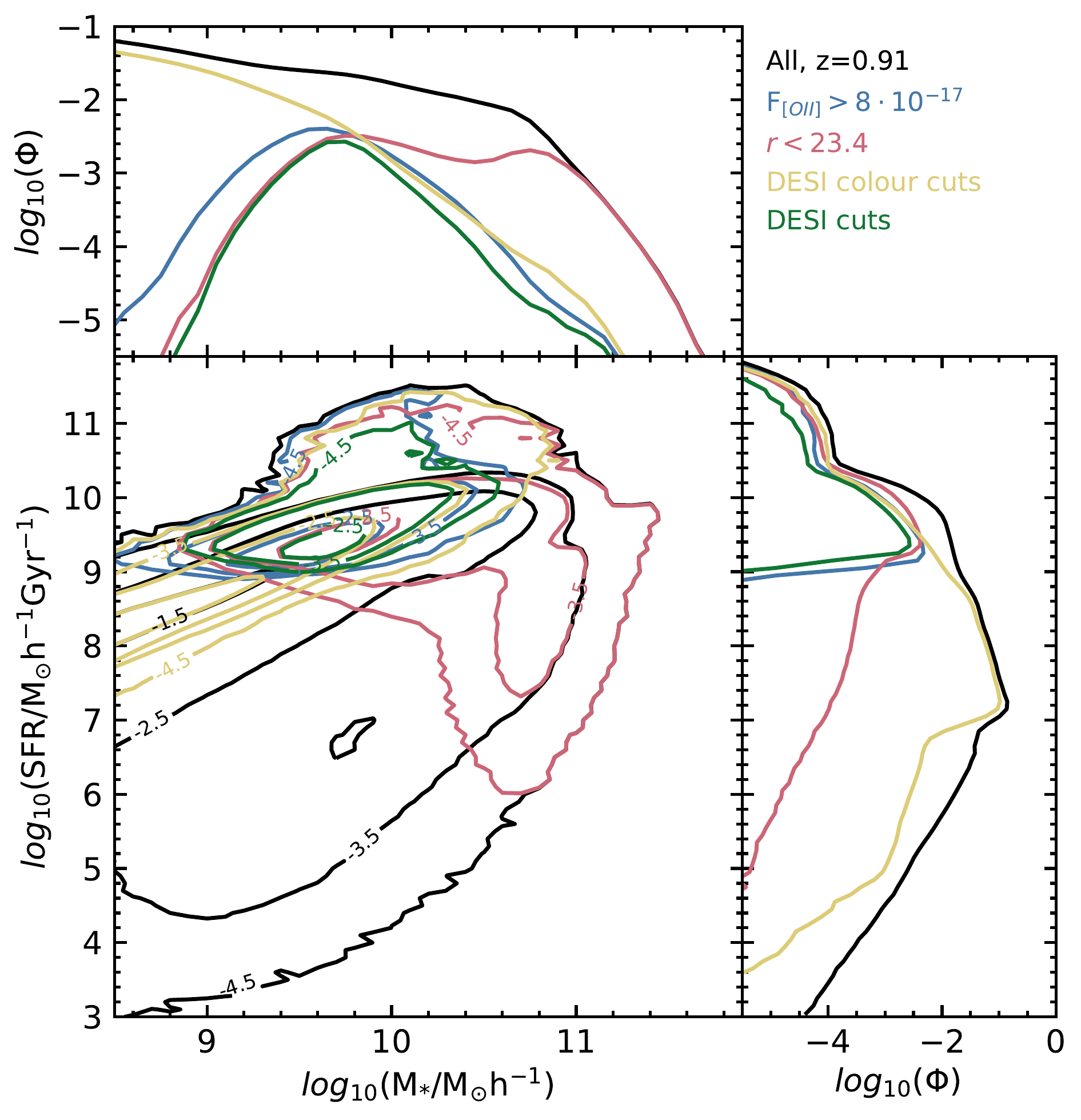}
\caption{\label{fig:gsmf_desi} Similar to Fig.~\ref{fig:gsmf_all} but illustrating the separated effects of the different DESI cuts. All model galaxies at $z=0.91$ are shown in black. Lines of the colours indicated in the legend show galaxies selected with $F_{[OII]}>8 \times 10^{17} {\rm erg \,s^{-1}\, cm^{-2}}$, $r<23.4$, only the DESI colour cuts and the full DESI cuts as summarised in Table~\ref{tbl:galform}. 
} 
\end{figure}

The \oem selected with the cuts presented in Table~\ref{tbl:obs} are star-forming galaxies \citep[e.g.][]{kewley04,moustakas06,mostek12}. Fig.~\ref{fig:gsmf_all} shows this by presenting the GP17 model SFR-stellar mass plane for all galaxies at $z=0.76$ and \oem selected by four of the cuts summarized in Table~\ref{tbl:obs}. Similar trends are found over the redshift range 0.6$\leq z \leq$1.5, whenever a sufficiently high galaxy number density is used. Most selections are dominated by the cut in line flux. For the mass resolution of our model, in the case of the VVDS-Deep and DEEP2 cuts, the fraction of star forming galaxies classified as \oem, varies by less than 2\% when only the cut in flux is applied.  The eBOSS and the DESI selections remove the brightest, $L{\rm [OII]}>10^{42}  h^{-2} {\rm erg\,s^{-1}}$, and the most strongly star-forming galaxies with their \oem selection criteria, as seen in Figs.~\ref{fig:lfall} and~\ref{fig:gsmf_all}. 

The effect of simply imposing a cut in flux in the SFR-stellar mass plane can be seen in Fig.~\ref{fig:gsmf_desi}. We find a clear correlation between the \ox luminosity and the average SFR such that a cut in \ox luminosity is approximately equivalent to selecting galaxies with a minimum SFR. This is with the exception of the most massive galaxies, which are removed when imposing a cut in \ox luminosity (as shown in the top panel of Figs.~\ref{fig:lf} and~\ref{fig:gsmf_desi}). Indeed, the most massive galaxies in the model are also those most affected by dust attenuation, in agreement with observational expectations \citep{sobral16}. 

The ELG samples summarised in Table~\ref{tbl:obs} are limited by their optical apparent magnitudes. Figs.~\ref{fig:gsmf_all} and~\ref{fig:gsmf_desi} show how this cut in apparent magnitude further reduces the number of low mass galaxies, with respect to a cut only in the \ox luminosity. Brighter galaxies in the optical tend to be brighter \oem and thus galaxies with either low masses or low SFR tend to be removed with a cut in apparent optical magnitude.

The distribution of model optical colours remains rather flat with \ox luminosity, and imposing the eBOSS colour cuts at a given redshift reduces the number of selected galaxies in a non-trivial way within the sSFR-stellar mass parameter space. We remind the reader that some of these colours cuts are actually imposed to remove low redshift galaxies, including ELGs, but also to avoid stellar contamination, as described in Appendix~\ref{sec:colours}.

Further ELG selection characteristics include: (i) the DEEP2 and VVDS-Deep cuts select over 95\% of \oem that form stars quiescently; (ii) galaxies with disks with radii greater than $3 h^{-1} {\rm kpc}$ account for $\sim$ 60\% of the \oem; (iii) the VVDS-Wide cut selects the brightest model \oem at the highest redshifts, while the eBOSS and DESI cuts remove the brightest model \oem with $L{\rm [OII]}>10^{42} h^{-2} {\rm erg\,s^{-1}}$. We note also that the brightest \ox emitters are dominated by spheroids that experience a burst of star formation. 

Given the above, a rough approximation to select \oem samples at $z\sim 1$ is to impose a cut in stellar mass, typically $M_{S}<10^{11} h^{-1} M_{\odot}$ (since massive galaxies in the model tend to be too attenuated and thus too faint to be selected as \oem) and SFR (which is tightly correlated to the cut in \ox luminosity), at least ${\rm SFR} >1 h^{-1} {\rm M_{\odot} yr^{-1}}$.


\section{The host haloes of \oem}
\begin{table}
\caption{The logarithm in base 10 of the mean mass (in units of $M_{\odot}h^{-1}$) of the haloes hosting the model \oem for the selections presented in Table~\ref{tbl:obs}. Values are shown at $z=0.76,\, 0.91,\, 1.17$ and $1.50$ for those selected \oem with a global de density above $10^{-4}{\rm Mpc}^{-3}h^{3}$. 
}
\hspace*{-0.7cm}
\label{tbl:mmhalo}
\begin{center}
\begin{tabular}{|c|c|c|c|c|}
\hline
Selection & $z=0.76$ & $z=0.91$ & $z=1.17$ & $z=1.50$ \\
\hline
   DEEP2 & ${11.41}$ & ${11.49}$ & ${11.55}$ & ${11.61}$\\
   VVDS-Deep & ${11.49}$ & ${11.54}$ & ${11.58}$ & ${11.64}$\\
   VVDS-Wide & ${11.71}$  &  ${11.78}$ & - & - \\
   eBOSS   & ${11.65}$ & ${11.74}$ & - & - \\
   DESI & ${11.46}$ & ${11.56}$ & ${11.63}$ & -\\
\hline
\end{tabular}
\end{center}
\end{table}
%

Model \oem at $0.5< z<1.5$ are hosted by haloes with masses above $10^{10.3}h^{-1}M_{\odot}$ and mean masses in the range $10^{11.41}\leq M_{\rm halo}(h^{-1}M_{\odot})\leq 10^{11.78}$, as summarised in Table~\ref{tbl:mmhalo}. From this table  it is clear that the host halo mean masses slightly increase with redshift in the studied range. These model masses are consistent with the estimation from \citet{khostovan17} for \oem at $z=1.47$.

In this section, the masses of haloes hosting \oem selected as indicated in Table~\ref{tbl:obs} are furthered explored through the stellar mass-halo mass relation and mean halo occupation distribution. 

\subsection{The stellar mass-halo mass relation}
\begin{figure}  
\includegraphics[width=0.5\textwidth]{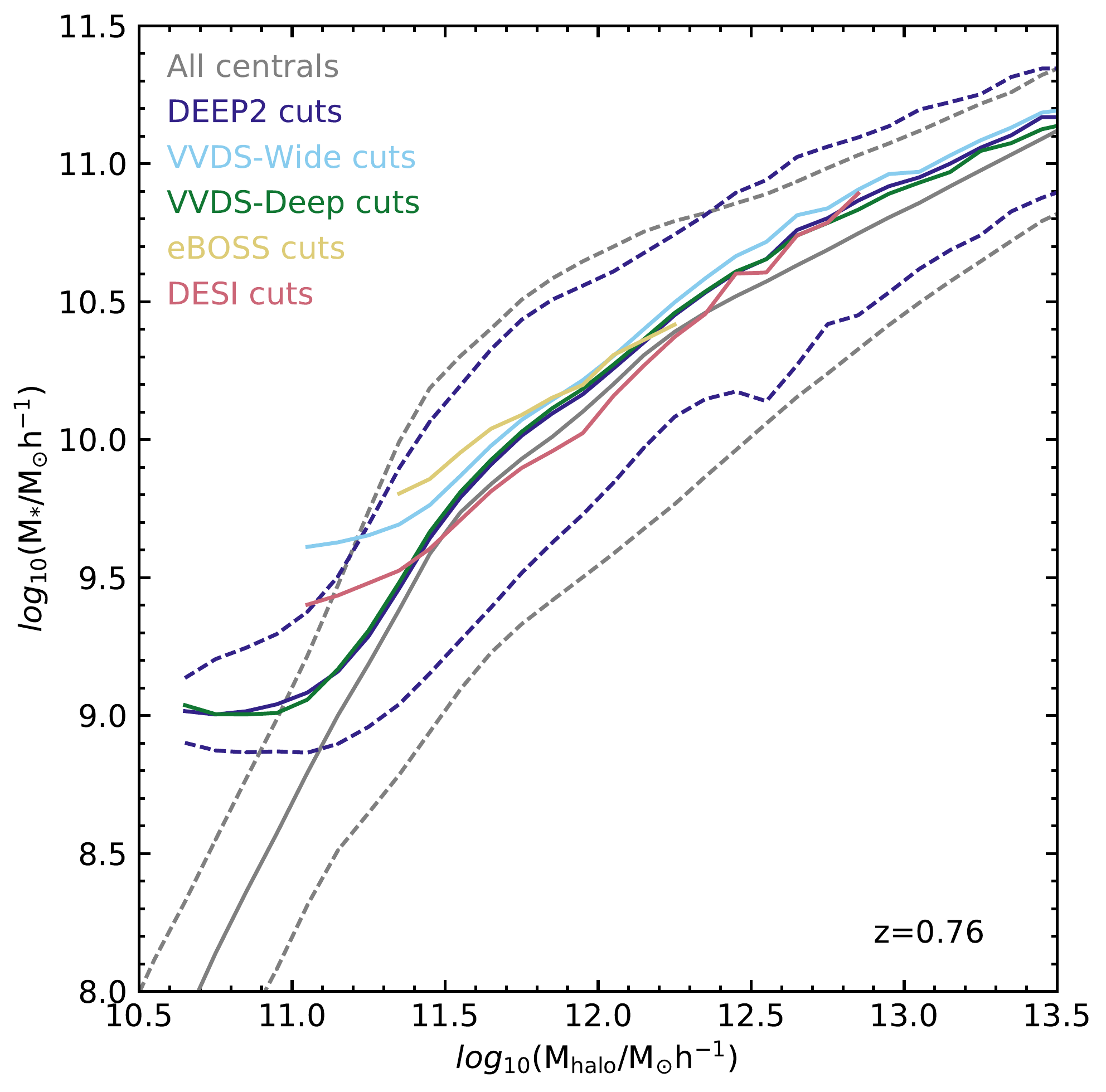}
\caption{\label{fig:mh_sm} The median stellar-to-halo mass relation for central galaxies in the GP17 model at $z=0.76$ (grey solid lines), with the 10$^{\rm th}$ and 90$^{\rm th}$ percentiles (grey dashed lines). 
The median relations for model central galaxies selected with specific survey cuts (see Table~\ref{tbl:obs}) are shown by the solid lines, colour coded following the key. For clarity, the 10$^{\rm th}$ and 90$^{\rm th}$ percentiles are showed only for the DEEP2 selection cut, and only halo mass bins with at least 100 galaxies are plotted.
} 
\end{figure}
The stellar-to-halo mass relation for model \oem is presented in Fig.~\ref{fig:mh_sm} at $z=0.76$, together with the global relation for central galaxies. We only show central galaxies in this plot as the sub-haloes hosting satellite galaxies are being disrupted due to tidal stripping and dynamical friction.

At low halo masses, the stellar-to-halo mass relation for the model \oem flattens out as the cut in the emission line flux effectively imposes a lower limit on the stellar mass of the selected galaxies (see Fig.~\ref{fig:gsmf_desi}). Above this flattening the stellar mass of model galaxies increases with their host halo mass, with a change of slope around ${\rm M_{halo}}\sim 10^{12} h^{-1} {\rm M}_{\odot}$, where star formation is most efficient at this redshift \citep[e.g.][]{behroozi10,moster13,rodriguezpuebla17}. At this halo mass the dispersion in the stellar-to-halo mass relation increases, being about 1.1 dex for all centrals in the model and between 0.5 and 0.8 dex for central \oem. This is a behaviour particular to \gl and it is related to the modelling of the growth of bulges \citep[see][for a more detailed discussion]{g16,mitchell16}.

For haloes with ${\rm M_{halo}}\sim 10^{12.5} h^{-1} {\rm M}_{\odot}$, the median stellar mass of model \oem  is $\sim1.5$ greater than that of the global population. This is driven by the cut in \ox flux removing low mass galaxies. The selection of \oem removes the most massive star forming galaxies because they are dusty on average and thus, the difference with respect to the global population is smeared out.

\subsection{The mean halo occupation distribution}\label{sec:hod}
\begin{figure}
\includegraphics[width=0.5\textwidth]{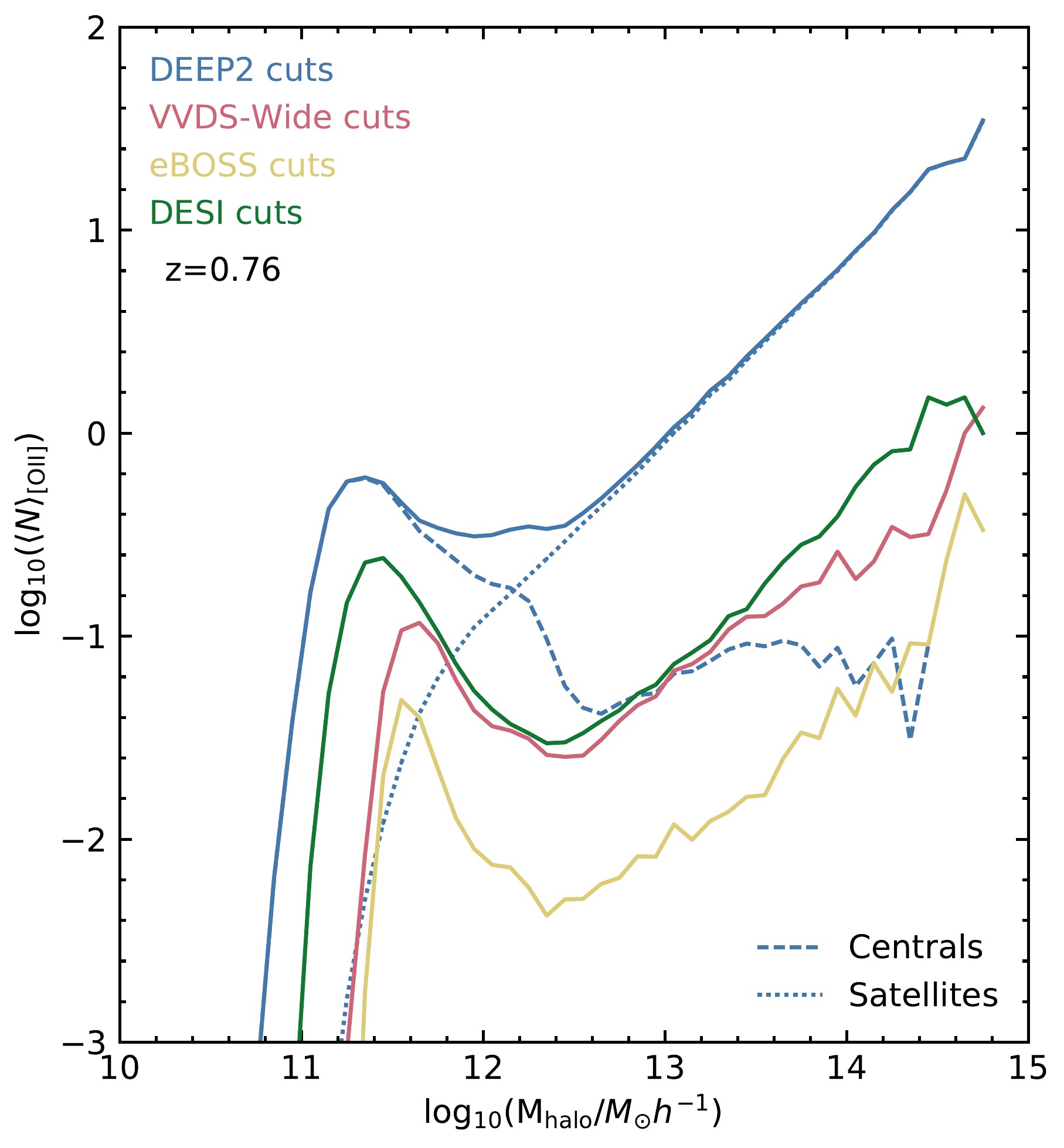}
\caption{\label{fig:hod} The mean halo occupation distribution, \nm (solid lines), for galaxies at $z=0.76$ selected using the cuts indicated in the legend (see Table~\ref{tbl:obs} for their definitions). For galaxies selected using the DEEP2 cuts, the contributions from central and satellite galaxies are shown as dashed and dotted lines respectively.} 
\end{figure}

\begin{figure}  \includegraphics[width=0.5\textwidth]{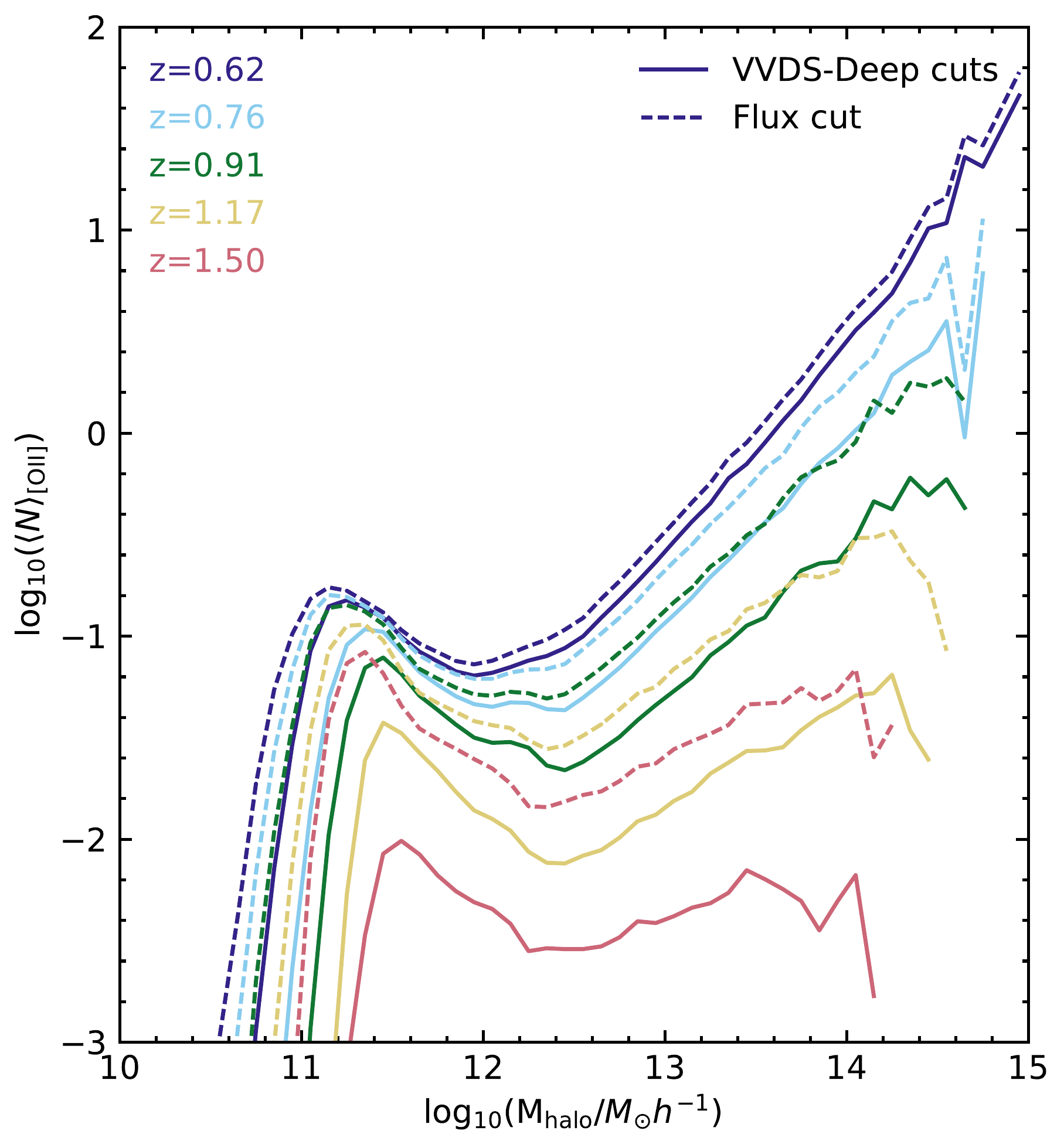}
\caption{\label{fig:hod_z} The redshift evolution (colour coded according to the legend, covering the range $z=0.62$ to $z=1.5$ from top to bottom) of the mean halo occupation distribution of \oem (\nmo) selected with the VVDS-Deep cuts (solid lines) and with only a flux cut of $ F_{[OII]}>1.9 \times 10^{-17} {\rm erg\, s^{-1}cm^{-2}}$ (dashed lines).} 
\end{figure}

The mean halo occupation distribution, \nm, encapsulates the average number of a given type of galaxy hosted by haloes within a certain mass range. \nm is usually parametrised separately for central, $\langle N \rangle_{\rm cen}$, and satellite galaxies, $\langle N \rangle_{\rm sat}$. When galaxies are selected by their luminosity or stellar mass, $\langle N \rangle_{\rm cen}$ can be approximately described as a smooth step function that reaches unity for massive enough host haloes, while $\langle N \rangle_{\rm sat}$ is close to a power law \citep[e.g.][]{berlind03,zheng05}. However, when galaxies are selected by their star formation rates, $\langle N \rangle_{\rm cen}$ does not necessarily reach unity \citep[e.g.][]{zheng05,contreras15,cowley16}. This implies that haloes above a certain mass will not necessarily harbour a star forming galaxy or, in our case, an ELG. For star forming galaxies, the shape of the $\langle N \rangle_{\rm cen}$ as a function of halo mass can also be very different from a step function and in some cases it can be closer to a Gaussian \citep[e.g.][]{geach12,contreras13}. 

Fig.~\ref{fig:hod} shows the \nm for model \oem, \nmo, selected following the specific survey cuts detailed in Table~\ref{tbl:obs}. \nmo does not reach unity for all the survey selections in the explored redshift range (see also Fig.~\ref{fig:hod_z}). This result is fundamental for interpreting the observed clustering of ELGs, as the standard expectation for $\langle N \rangle_{\rm cen}$ is to tend to unity for large halo masses. This point is further emphasized by splitting \nmo for galaxies selected with DEEP2-like cuts into satellites and  centrals. The \nm of model central \oem, \nmoc, is very different from the canonical smooth step function, which is usually adequate to describe  stellar mass threshold samples and is the basis of (sub) halo abundance matching \citep[e.g.][]{vale04,conroy06}.
We further discuss the \nmoc in \S~\ref{sec:centrals}.

On the other hand the predicted $\langle N \rangle_{\rm sat}$ of \oem closely follows the canonical power law above a minimum halo mass that is typically an order of magnitude larger than the minimum halo mass required to host a central galaxy with the same selection. In the cases studied, less than 10\% of the modelled \oem are satellite galaxies, and thus there are very few haloes hosting even one satellite \ox emitter.

The redshift evolution of the \nmo is presented in Fig.~\ref{fig:hod_z} for both galaxies selected with the VVDS-Deep cuts and for a simple cut in the \ox flux line, $F_{[OII]}>1.9\times 10^{-17} {\rm erg\, s^{-1}cm^{-2}}$. There is a clear drop with redshift for all halo masses in the average halo occupancy of model \oem. This is mostly driven by the survey magnitude cut, as a simple flux cut reduces the mean occupation much more gradually with redshift. Over the redshift range probed, the minimum mass of haloes hosting \oem increases with redshift, as they are selected for a fixed cut in either \ox flux line alone or also in apparent magnitude. 

Finally we note that similar \nmo shapes are seen for the other cuts considered in this redshift range, with the main change being in the average number of galaxies occupying a given mass halo.

\subsection{\ox central galaxies}\label{sec:centrals}
\begin{figure}  \includegraphics[width=0.5\textwidth]{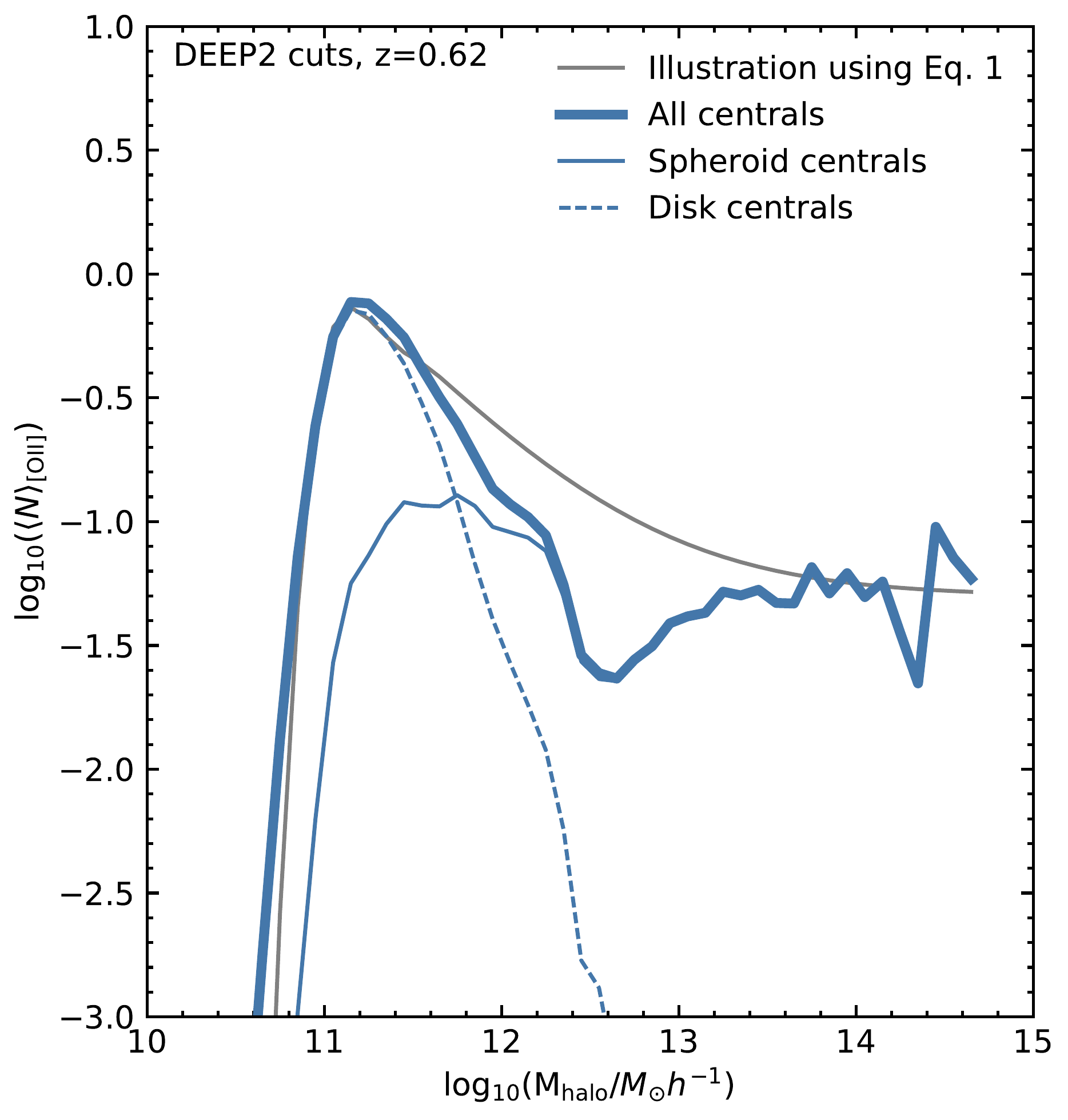}
\caption{\label{fig:hod_bot} The \nm for central \oem selected with the DEEP2 cuts at $z=0.62$ (thick line). The $\langle N \rangle_{\rm cen}$ is split into the contribution from spheroid-dominated (solid) and disk-dominated (dashed) galaxies. The latter correspond to galaxies with a bulge over total mass less or equal to 0.5. An illustration of Eq.~\ref{eq:hod} is shown in grey (see \S~\ref{sec:centrals}). } 
\end{figure}

As seen in Fig.~\ref{fig:hod} already, the \nmoc is clearly different from a step function. Note that this shape cannot be recovered if a cut in SFR and stellar mass is applied, similar to the rough approximation to select \oem suggested at the end of \S~\ref{sec:exploring}. The shape of the \nmoc seen in Fig.~\ref{fig:hod} for model central \oem is closer to a Gaussian plus a step function or even a power law. This might point to the contribution of at least two different types of model central \oem. We have explored splitting central \oem in different ways, including separating those experiencing a burst of star formation. When splitting central \oem into disks and spheroid galaxies, using a bulge over total mass ratio of 0.5 to set the disk-spheroid boundary, we recover an \nmoc that can be roughly described as an asymmetric Gaussian, for disk centrals, plus a step function that rises slowly to a plateau, for bulges or spheroid centrals. This is shown in Fig.~\ref{fig:hod_bot} for \oem selected with DEEP2 cuts at $z=0.62$, but similar results are found for other selections and redshifts, as long as the number density of galaxies is sufficiently large for the split to remain meaningful. 

Surveys such as eBOSS and DESI will obtain low resolution spectra for \oem which are unlikely to be sufficient to gather the information needed to split the population into disks and spheroids. Within the studied redshift range, model \oem that are central disks, tend to be less massive, have lower stellar metallicities and larger sizes than central spheroids, for all the selections presented in Table~\ref{tbl:obs}. In particular, for a given halo mass central galaxies that are spheroids have stellar masses up to a factor of 1.6 larger than central discs. However, since the bulge to total mass ratio varies smoothly with stellar mass, the distributions of these model properties have a large overlap for central disks and spheroids and thus, it is unclear if they could be used observationally to split the central \oem population. 

A split into three components might describe better the \nmoc presented in Fig.~\ref{fig:hod_bot}. However, on top of the \nm becoming noisy for large halo masses it will already be difficult to split observed central \oem into disks and spheroids to test our model, as in most cases only spectroscopic information is available. Thus, to encapsulate into an illustrative function the shape of the \nm for model central \oem, we have opted to propose a function that adds together a softly rising step function for central spheroids (or bulges), {\it b}, with an asymmetric Gaussian for central disks, {\it d}:

\begin{equation}\label{eq:hod}
\begin{split}
\langle N \rangle_{\rm cen} &= \frac{f_b}{2} \left( 1 + {\rm erf} \left( \frac{{\rm log}_{10}{\rm M}_{*}- {\rm log}_{10}{\rm M}_b}{\sigma} \right) \right) + \\
& \frac{f_d}{2}e^{\frac{\alpha_d}{2}\left( 2{\rm log}_{10}{\rm M}_d + \alpha_d\sigma^2 - 2{\rm log}_{10}{\rm M}_* \right)}  \times \\
&{\rm erfc} \left( {\rm log}_{10}{\rm M}_d + \alpha_d\sigma ^2 - \frac{{\rm log}_{10}{\rm M}_*}{\sigma\sqrt{2}} \right)
\end{split}
\end{equation}

In the above equation, ${\rm erf}$ is the error function (${\rm erfc}=1-{\rm erf}$)\footnote{${\rm erf}(x)=\frac{2}{\sqrt[]{\pi}}\int^x_0e^{-t^2}dt$}, which behaves like a softly rising step function. $M_b$ gives the characteristic halo mass of the error function for the central bulges, and $M_d$ gives the average halo mass of the Gaussian component for central disks. $f_b$ and $f_d$ control the normalisation of the error function and the Gaussian component, respectively. $\sigma$ controls the rise of the error function and the width of the asymmetric Gaussian. The level of asymmetry of the Gaussian component is controlled both by $\sigma$ and $\alpha_d$. 

As an illustration, Fig.~\ref{fig:hod_bot} shows in grey the function described in Eq.~\ref{eq:hod} with parameters: ${\rm log}_{10}M_b=11.5$, ${\rm log}_{10}M_d=11.0$, $f_b=0.05$, $f_d=1$, $\sigma=0.09$, $\alpha _d=1.7$. Adequately fitting the shape of the \nmoc with Eq.~\ref{eq:hod} is out of the scope of this paper. Moreover, an individual fit to disk and spheroid central galaxies will be more adequate. We defer such an exploration because, as it will be discussed in the next section \S~\ref{sec:xi}, it is unclear that our model is producing a large enough number of \oem that are also satellite galaxies compared to the expectations from observations. Moreover, the proposed split might not actually be achieved observationally. Nevertheless, given that uncommon features in the mean HOD can affect the inferred galaxy clustering \citep[][]{nuala17}, our proposed Eq.~\ref{eq:hod} is a useful tool to explore the impact that such a mean HOD has when interpreting mock catalogues generated for cosmological purposes.

\section{The clustering of \oem}\label{sec:xi}
\begin{figure} 
\includegraphics[clip, trim=2.2cm 0.2cm 2.5cm 0.5cm, width=.5\textwidth]{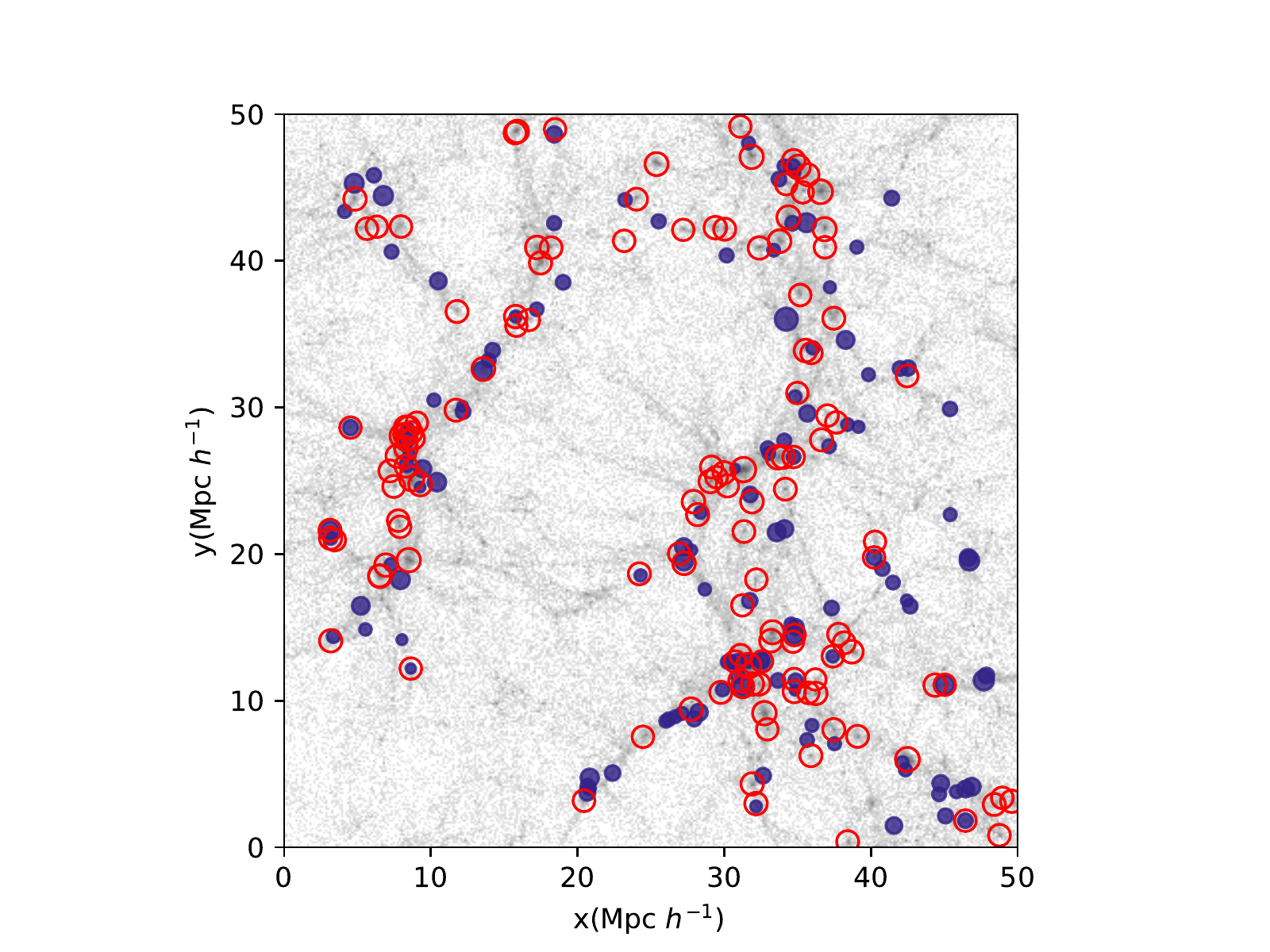}
\caption{\label{fig:pretty} Distribution of \oem selected with the DEEP2 cuts (filled blue circles) and that of dark matter haloes above $10^{11.8}h^{-1}$M$_{\odot}$ (open red circles) painted on top of the smooth underlying dark matter distribution (grey). This slice of 10 Mpc$/h$ thickness is taken from the MS-W7 simulation at $z=1$. The halo mass cut is defined so as to match the number density of model \oem galaxies, i.e.\ $0.005 h^{3}$Mpc$^{-3}$. The circles area is proportional to the log$_{10}(L_{[OII]})$ and the log$_{10}(M_h)$ for \oem and dark matter halos respectively.}
\end{figure}

In this section we explore how \oem trace the dark matter distribution. In Fig.~\ref{fig:pretty} we present a $50\times 50\times 10 \, h^{-3} \, {\rm Mpc}^3 $ slice of the whole simulation box at redshift $z=1$, highlighting in grey the cosmic web of the dark matter, together with the location of \oem (filled circles) and of dark matter haloes above $10^{11.8}h^{-1}$M$_{\odot}$ (open circles). The environment where model \oem are found is not the densest as expected for other cosmological tracers such as luminous red galaxies, but instead the \oem are also found in filamentary structures.

Below we explore the two point correlation function monopole in both real and redshift space for model \oem. The two point correlation function has been obtained using two algorithms that give similar results; the plots show the calculation from the publicly available {\tt CUTE} code \citep{cute}. The linear bias is also calculated in real space and we compare it with the expectations for eBOSS and DESI (\S~\ref{sec:xi_r}). \citet{favole16} measured the redshift space monopole for a sample at $z\sim 0.8$ of g-selected galaxies that they claim is comparable to a selection of \oem. We also make cuts similar to those in \citeauthor{favole16} in order to compare the results for g-band selected galaxies and \oem with their observed clustering (\S~\ref{sec:xi_z}).

\subsection{The correlation function and galaxy bias in real-space}\label{sec:xi_r}
\begin{figure} 
\includegraphics[width=.5\textwidth]{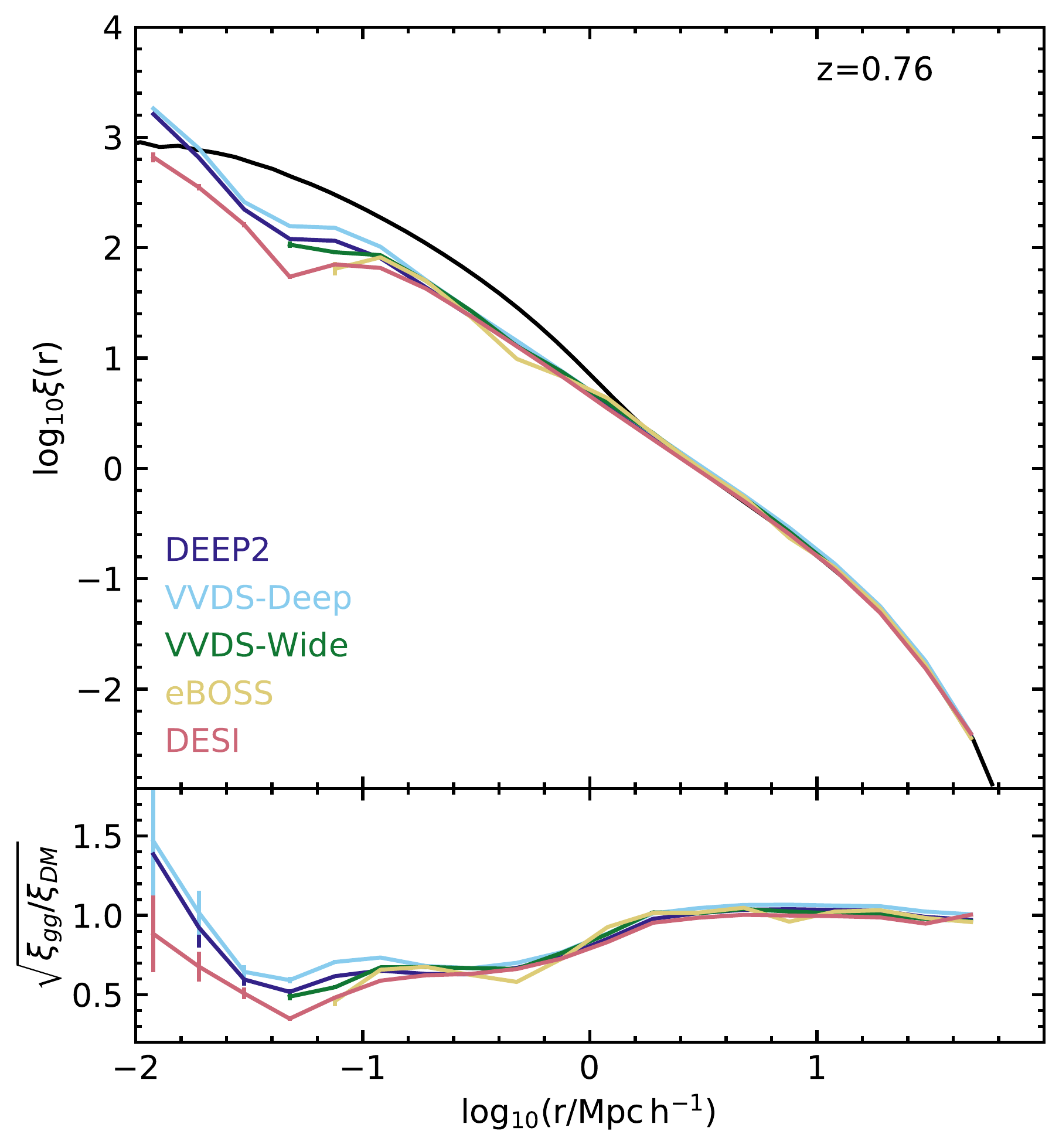}
\caption{\label{fig:xi_surveys} {\it Top panel:} The real-space two point correlation function at $z=0.76$ for model galaxies from each of the selections indicated in the legend (see Table~\ref{tbl:obs}) together with that of the underlying dark matter (black line). {\it Bottom panel:} The real space bias, $\sqrt{\xi_{gg}/\xi_{DM}}$, at the same redshift. Poisson errors are shown in both panels.}
\end{figure}

\begin{figure} 
\includegraphics[width=.5\textwidth]{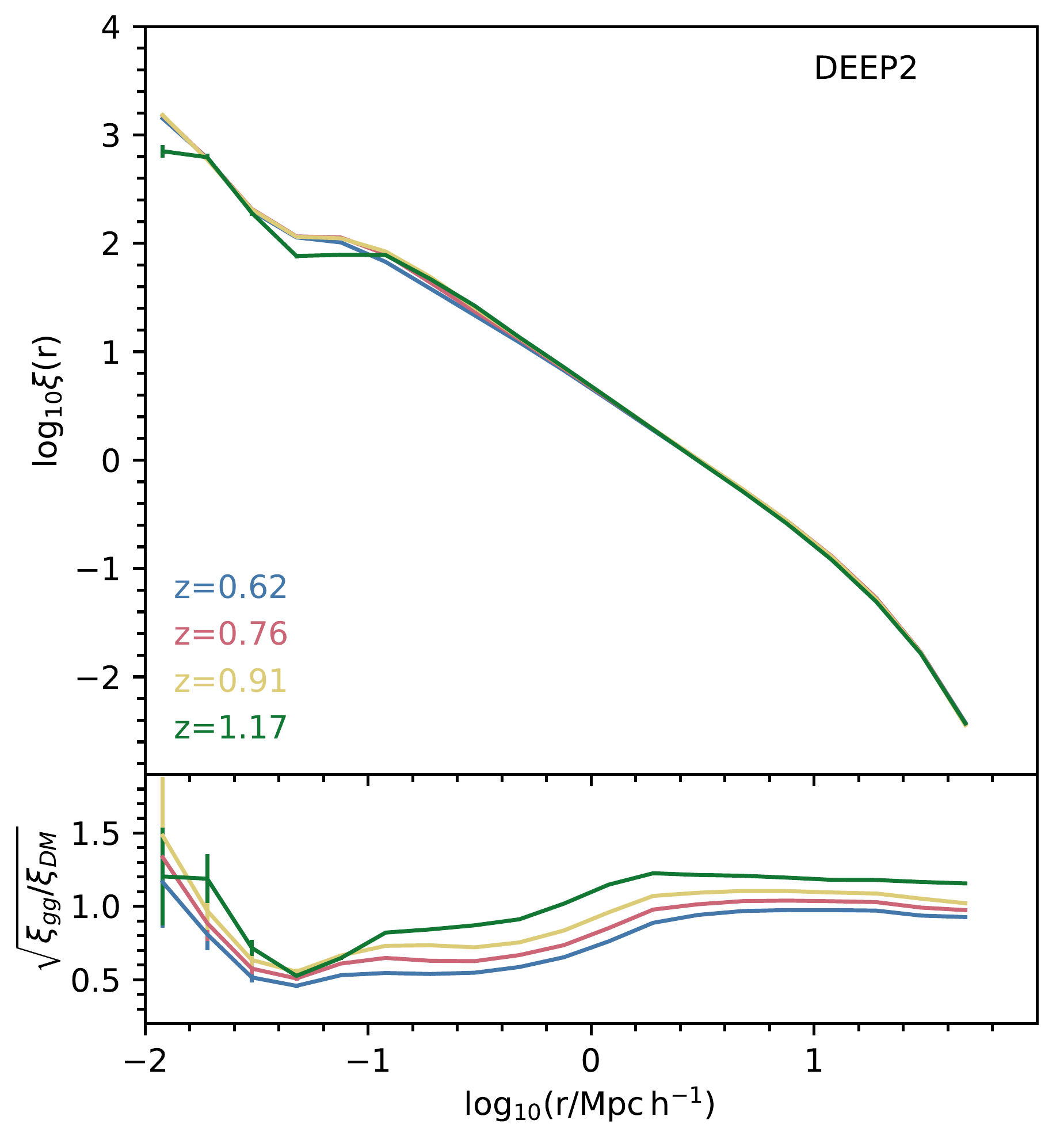}
\caption{\label{fig:xi_zs} {\it Top panel:} The real-space two point correlation function for model galaxies selected with the DEEP2 cuts (see Table~\ref{tbl:obs}) at the redshifts indicated in the legend. {\it Bottom panel:} The real space bias, $\sqrt{\xi_{gg}/\xi_{DM}}$, at the same redshifts. Poisson error bars are shown in both panels.}
\end{figure}

Fig.~\ref{fig:xi_surveys} shows the real-space two point correlation function, $\xi(r)$, for model galaxies at $z=0.76$ selected following the cuts in Table~\ref{tbl:obs}. The different galaxy selections result in a very similar $\xi(r)$, in particular on scales above $0.1 h^{-1}$Mpc. The same is true for the other redshifts explored. Compared to the dark matter real-space two point correlation function, $\xi_{\rm DM}$, model galaxies follow closely the dark matter clustering for comoving separations greater than $\sim1 h^{-1}$Mpc. The real space bias, $\sqrt{\xi_{gg}/\xi_{DM}}$, is practically unity and constant for comoving separations greater than $2 h^{-1}$Mpc. In comparison, SDSS luminous red galaxies (LRGs) have a bias of $\sim 1.7\,\sigma_8(0)/\sigma_8(z)$\footnote{$\sigma_8(z)$ gives the normalization of the density fluctuations in linear theory and has a value of 0.81 at $z=0$ for the cosmology assumed in this work.}. From a pilot study, eBOSS ELGs are expected to be linearly biased, with a bias of $\sim 1.0\,\sigma_8(0)/\sigma_8(z)$ \citep{dawson16}. For the cosmology assumed in this study, eBOSS LRGs are then expected to have a bias of 2.7 at $z=1$ and ELGS have $b=1.62$ at the same redshift\footnote{The ratio $\sigma_8(0)/\sigma_8(z)$ has been obtained using the ICRAR cosmological calculator \url{http://cosmocalc.icrar.org/}.}.

Fig.~\ref{fig:xi_zs} shows the evolution of the bias over the redshift range of interest for this study for DEEP2 model galaxies. For both DEEP2 and VVDS-Deep selections, the bias on large scales increases by a factor of 1.2 from $z=0.6$  to $z=1.2$. For all the considered selections, when the propagated Poisson errors are below $\sigma_b=0.2$, the linear bias remains between 1 and 1.4 in all cases. 

Given the predicted small fraction of \oem that are satellite galaxies, these galaxies have the potential to be extraordinary cosmological probes for redshift space distortion analysis as they are possibly almost linearly biased for the 2-halo term.

\subsection{The correlation function in redshift-space}\label{sec:xi_z}
\begin{figure} 
\includegraphics[width=.5\textwidth]{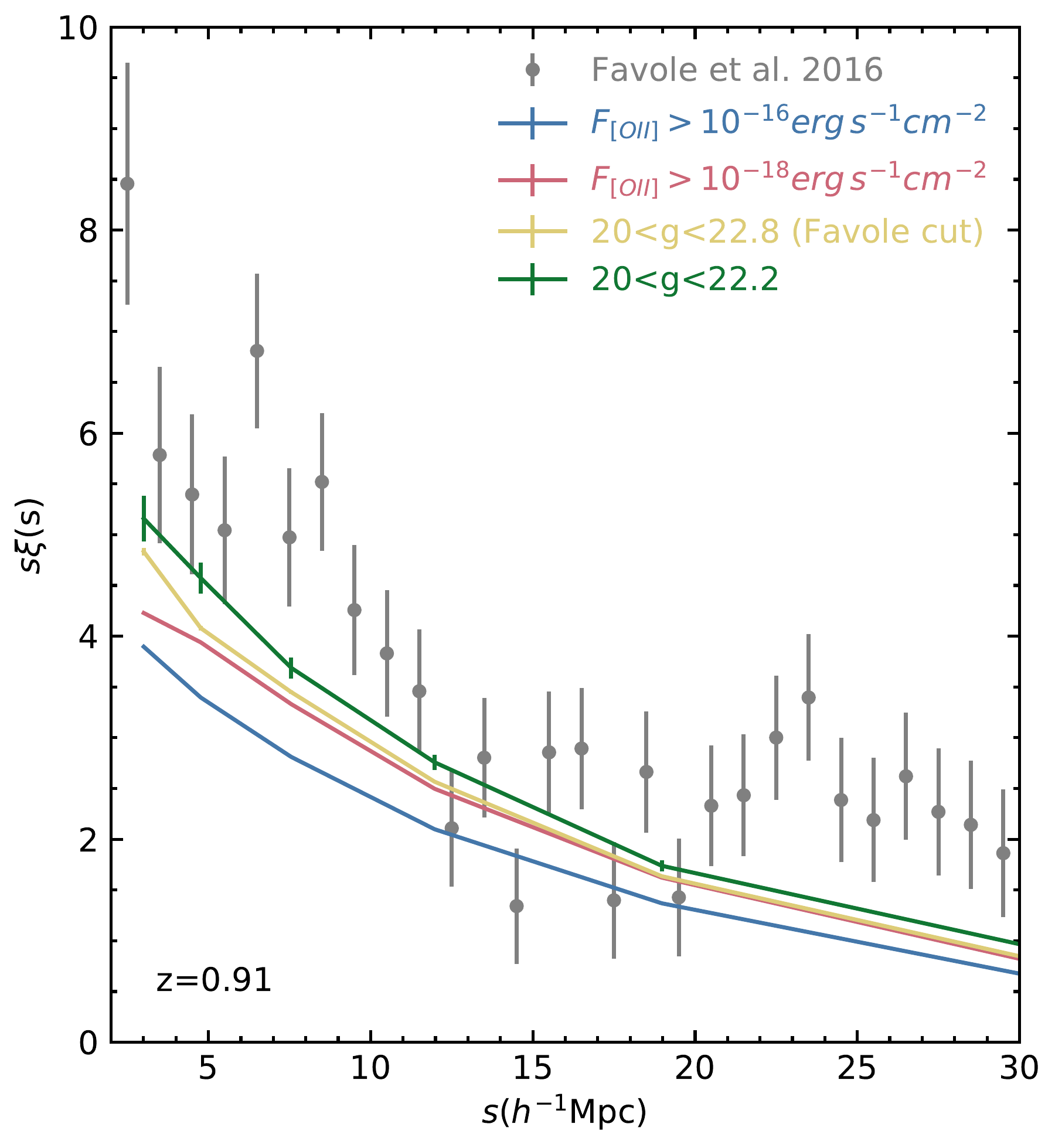}
\caption{\label{fig:xi_s} The predicted redshift-space two point correlation function, $\xi(s)$, at $z=0.91$ plotted scaled by the comoving separation, $s$, as a function of $s$, for \ox flux and g-band selected galaxies as indicated in the legend. Predictions for model galaxies are shown with Poisson error bars. The grey symbols show the observational results presented in \citet{favole16}, with error bars that include sample variance.}
\end{figure}

The redshift-space two point correlation function is shown in Fig.~\ref{fig:xi_s} for model galaxies selected with two different \ox flux and g-band cuts. Model galaxies with a brighter [OII] flux are less clustered in both real and redshift space. This contradicts what is found observationally at $z=0$ \citep{favole16} and is related to the number of star-forming galaxies satellites in the model (see Fig.~\ref{fig:pf_cal}). Samples of model \oem are hosted by haloes with minimum masses that increase with the \ox flux. However, the fraction of satellite \oem decreases for brighter cuts in \ox flux. At $z=0.91$, the percentage of satellites is reduced from 20\% to 4\% when the selection in \ox flux is changed from $10^{-18} {\rm erg\, s^{-1}cm^{-2}}$ to $10^{-16}{\rm erg\, s^{-1}cm^{-2}}$. This reduction in the numbers  of model satellite galaxies in bright samples of \oem, lowers the average mass of their host haloes, reducing the bias and clustering predicted by the model. 

\citet{favole16} measured the clustering of a g-band selected galaxy sample, with an average density of 500 galaxies per deg$^2$ in the redshift range $0.6<z<1.7$. The selection  of galaxies based on their apparent g-band magnitude around $z=1$ is very close to selecting ELGs. \citet{comparat15o2} showed that the g-band magnitude is correlated to the \ox luminosity in the studied redshift range. We also find such a correlation for \oem in the model. This correlation is due to the fact that emission lines are directly related to the rest-frame UV luminosity, as this gives a measure of the ionizing photons.

The two point correlation functions for galaxies with $20<g<22.8$ and a colour cut to remove low redshift galaxies as measured by \citeauthor{favole16} are shown with grey symbols in Fig.~\ref{fig:xi_s}. In this figure we compare model galaxies selected with the same g-band cut to the results from \citet{favole16}. Note that the clustering of model galaxies with $F_{[OII]}>10^{-18}erg\, s^{-1}cm^{-2}$ and $20<g<22.8$ overlap for separations above 10$h^{-1}$Mpc and below this separation they are comparable. The reduced $\chi^2$ is 3.1 when comparing the clustering of model galaxies with $20<g<22.8$ to that of \citet{favole16}. The reduced $\chi^2$ decreases to $\sim2.6$, if the g-band faint cut is changed by 0.6 magnitudes, $20<g<22.2$. Model galaxies appear to be less clustered than the current observations of g-selected samples. 

\citet{favole16} used weak lensing to estimate the typical mass of haloes hosting g-band selected galaxies, finding $(1.25\pm0.45)\times 10^{12}h^{-1}$M$_{\odot}$. Within the same redshift range, the model g-band sample is hosted by haloes  with an average mass of $\sim 10^{11.8}h^{-1}$M$_{\odot}$, consistent with the values reported observationally, although somewhat on the low side. 

\citet{favole16} also estimated the fraction of satellite $g$-band selected galaxies using a modified sub-halo abundance matching method that accounts for the incompleteness of small samples of galaxies that do not populate every halo. The model that best fits their measured clustering had $\sim20$\% satellite galaxies, while here we find that satellites account for only 2\% of our sample. 

Both aspects, the lower satellite fraction and slightly lower host halo masses contribute to explaining the lower two point correlation function obtained for model $g$-band selected galaxies in comparison to the observational results from \citeauthor{favole16} This result suggests that too large a fraction of model satellite galaxies are not forming stars at $z\sim 1$. In fact, even at $z=0$ we find too large a fraction of low mass galaxies with a very small star formation rate, compared to the observations (see Fig.~\ref{fig:pf_cal}, note that the problem is even larger for the GP14 model). The obvious place to start improving the model would be to allow satellite galaxies to retain their gas for longer, so they can have higher star formation rates on average. However, a thorough exploration of how expelled gas is reincorporated at different cosmic times might be needed \citep{mitchell14,henriques15,hirschmann16}.

\section{Conclusions}\label{sec:conclusions}
The GP17 semi-analytical model is a new hierarchical model of galaxy formation and evolution that incorporates the merger scheme described in \citet{simha16} and the gradual stripping of hot gas in merging satellite galaxies \citep{font08,lagos14.h1}. The GP17 model also includes a simple model for emission lines in star-forming galaxies that uses the number of ionizing photons and metallicity of a galaxy to predict emission line luminosities based on the properties of a typical HII region \citep{sta90}.

The free parameters in the GP17 model have been chosen to reproduce at $z=0$ the rest-frame luminosity functions (LF) in the $b_J$ and $K$ bands and also to improve the match to the local passive fraction of galaxies. 

Using the GP17 model, we study the properties of \oem. These are the dominant emission line galaxies (ELGs) selected by optical-based surveys at $0.5<z<1.5$. In particular, we have applied emission line flux, magnitude and colour cuts to the model galaxies, to mimic five observational surveys DEEP2, VVDS-Deep, VVDS-Wide, eBOSS and DESI, as summarised in Table~\ref{tbl:obs}. Over 99\% of the selected model \oem are actively forming stars, and over 90\% are central galaxies. 

The GP17 LFs of model \oem are in reasonable agreement with observations (see \S~\ref{sec:lf}). For this work, we have assumed that the dust attenuation experienced by the emission lines is the same as that for the stellar continuum. However, the assumed dust attenuation in the emission lines is expected to be a lower limit, which may alter the LF comparison. 

The bright end of the LF of \oem is dominated by galaxies undergoing a starburst. The luminosity at which this population dominates depends on the interplay between the stellar and the AGN feedback.

For model galaxies, we find that the cut in \ox luminosity removes galaxies below a certain SFR value, but that it also removes the most massive galaxies in the sample due to dust attenuation of the \ox line (see \S~\ref{sec:exploring}). 

Model \oem are typically hosted by haloes with masses above $10^{10.3}h^{-1}M_{\odot}$ and mean masses in the range $10^{11.41}\leq M_{\rm halo}(h^{-1}M_{\odot})\leq 10^{11.78}$ (see Table~\ref{tbl:mmhalo}). For haloes with M$_{halo}\sim 10^{12.5}M_{\odot}h^{-1}$, model \oem have median stellar masses a factor of 1.5 above the global population. This is driven by the cut on \ox luminosity being directly translated into a cut in SFR, which in turn is correlated with stellar mass and thus, low mass galaxies are also being removed from the selection.

As expected for star forming galaxies, the mean halo occupation of central \oem, \nmoc,cannot be described by a step function that reaches unity above a certain host halo mass (the typical shape for mass selected galaxies). The \nmoc can be approximately decomposed into an asymmetric Gaussian for central disk galaxies, i.e.\ with bulge-to-mass ratio below $0.5$, and a smoothly rising step function for central spheroids, which, in general, would not reach unity (see \S~\ref{sec:centrals}). This last point implies that not every dark matter halo is expected to host an ELG and it is particularly relevant for HOD models used to populate very large dark matter simulations with cosmological purposes.

Model \oem at $z \sim 1$ have a real-space two point correlation function that closely follows that of the underlying dark matter above separations of $1 h^{-1} \, {\rm Mpc}$, resulting in a linear bias close to unity. This is lower than the preliminary results for eBOSS ELGs, by a factor of $\sim1.6$ (see \S~\ref{sec:xi_r}).

We have compared the clustering of $g$-band selected model galaxies with the observational results from \citet{favole16}, who argue that the cut $20<g<22.8$ selects ELGs at $0.6<z<1$, once an additional colour cut is applied to remove lower redshift galaxies. The typical mass of haloes hosting such $g$-band selected galaxies as inferred from weak lensing in \citeauthor{favole16} is consistent with the values we find for our corresponding model galaxies (see \S~\ref{sec:xi_z}). However, our model g-selected galaxies are slightly less clustered in redshift space compared to the findings of \citet{favole16}. This is mostly due to the smaller fraction of $g$-band selected satellites in GP17, $\sim 2$\%, compared to their $\sim 20$\%. \citeauthor{favole16} inferred the satellite fraction from a modified sub-halo abundance matching model that accounts for incompleteness, as not all haloes above a certain mass contain a g-band selected galaxy. This is an indication that too large a fraction of model satellite galaxies are not forming stars at $z\sim 1$. This suggests that our model of galaxy formation and evolution can be improved by allowing satellite galaxies to retain their hot halo gas for longer, so their average star formation range is increased. However, other possibilities should be also explored, such as the reincorporation of expelled gas through cosmic time, which will most likely also have an impact on the selection of star forming satellite galaxies.

Future theoretical studies of emission line galaxies will benefit from the use of a more realistic model for the mechanisms that produce emission line galaxies. Given the small fraction of \oem that are satellite galaxies, ELGs have the potential to became ideal candidates for redshift space distortions studies at different cosmic times, due to the ease of modelling their clustering. However, the non-canonical shape of their mean halo occupation distribution should be studied and maybe accounted for in cosmological studies.

\section*{Acknowledgements}

The authors would like to thank the helpful comments provided by Will Percival, Will Cowley, Coleman Krawczyk, Carlos Alvarez, Marc Kassis, Zheng Zheng, Charles Jose and Anal\'ia Smith-Castelli. VGP acknowledges support from the University of Portsmouth through the Dennis Sciama Fellowship award and past support from a European Research Council Starting Grant (DEGAS-259586). JRH acknowledges support from the Royal Astronomical Society Grant for a summer studentship. VGP and JRH thank Alastair Edge for his help getting this studentship. PN, CMB, CL, NM and JH were supported by the Science and Technology Facilities Council (ST/L00075X/1). PN acknowledges the support of the Royal Society through the award of a University Research Fellowship, and the European Research Council, through receipt of a Starting Grant (DEGAS-259586). This work used the DiRAC Data Centric system at Durham University, operated by the Institute for Computational Cosmology on behalf of the STFC DiRAC HPC Facility (www.dirac.ac.uk). This equipment was funded by BIS National E-infrastructure capital grant ST/K00042X/1, STFC capital grants ST/H008519/1 and ST/K00087X/1, STFC DiRAC Operations grant ST/K003267/1 and Durham University. DiRAC is part of the National E-Infrastructure. This work has benefited from the publicly available programming language {\sc python}.




\bibliographystyle{mnras}
\bibliography{elg_hod.bbl} 



\appendix

\section{Colour cuts}\label{sec:colours}

\begin{figure}  \includegraphics[width=0.45\textwidth]{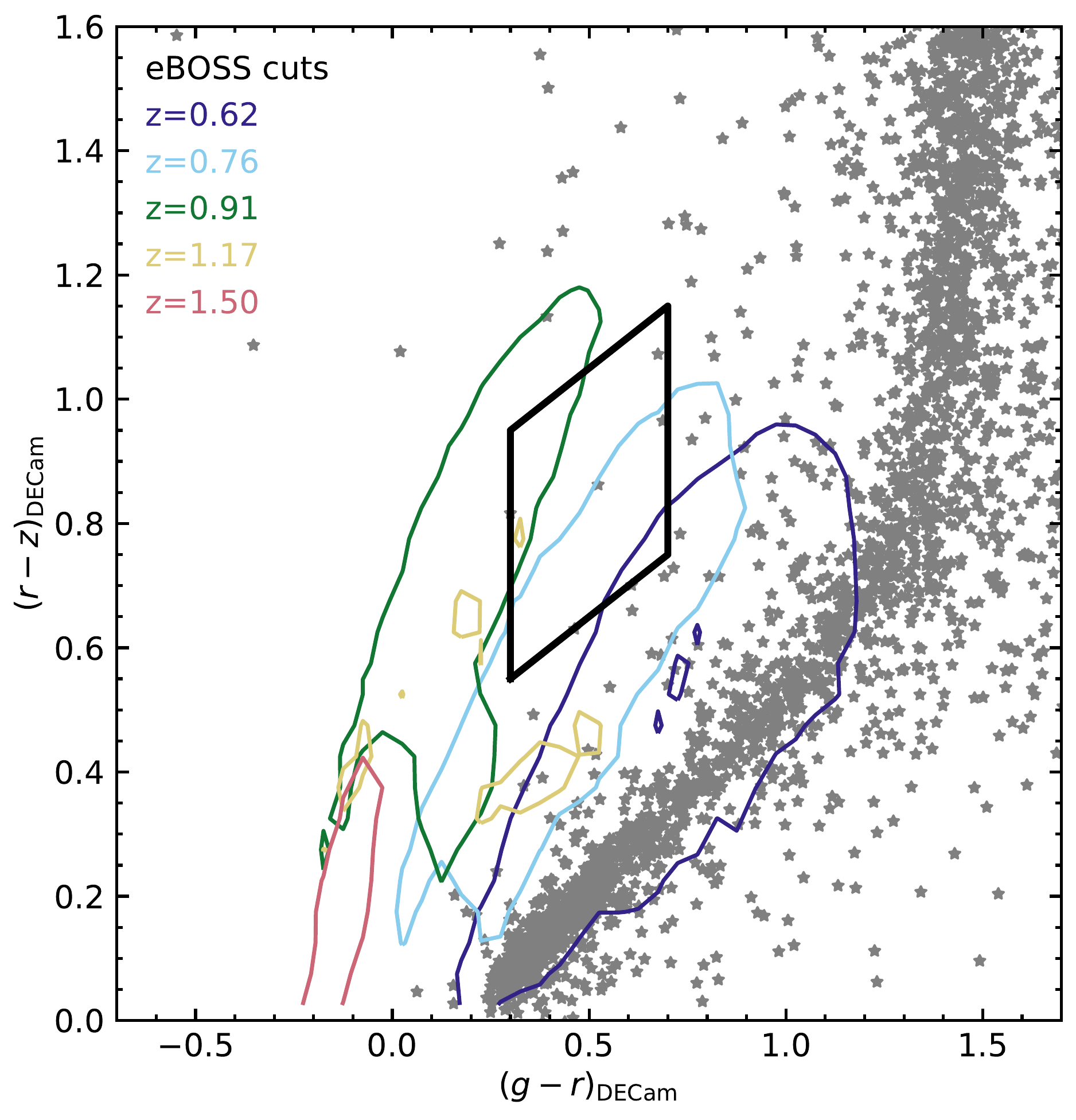}

\includegraphics[width=0.45\textwidth]{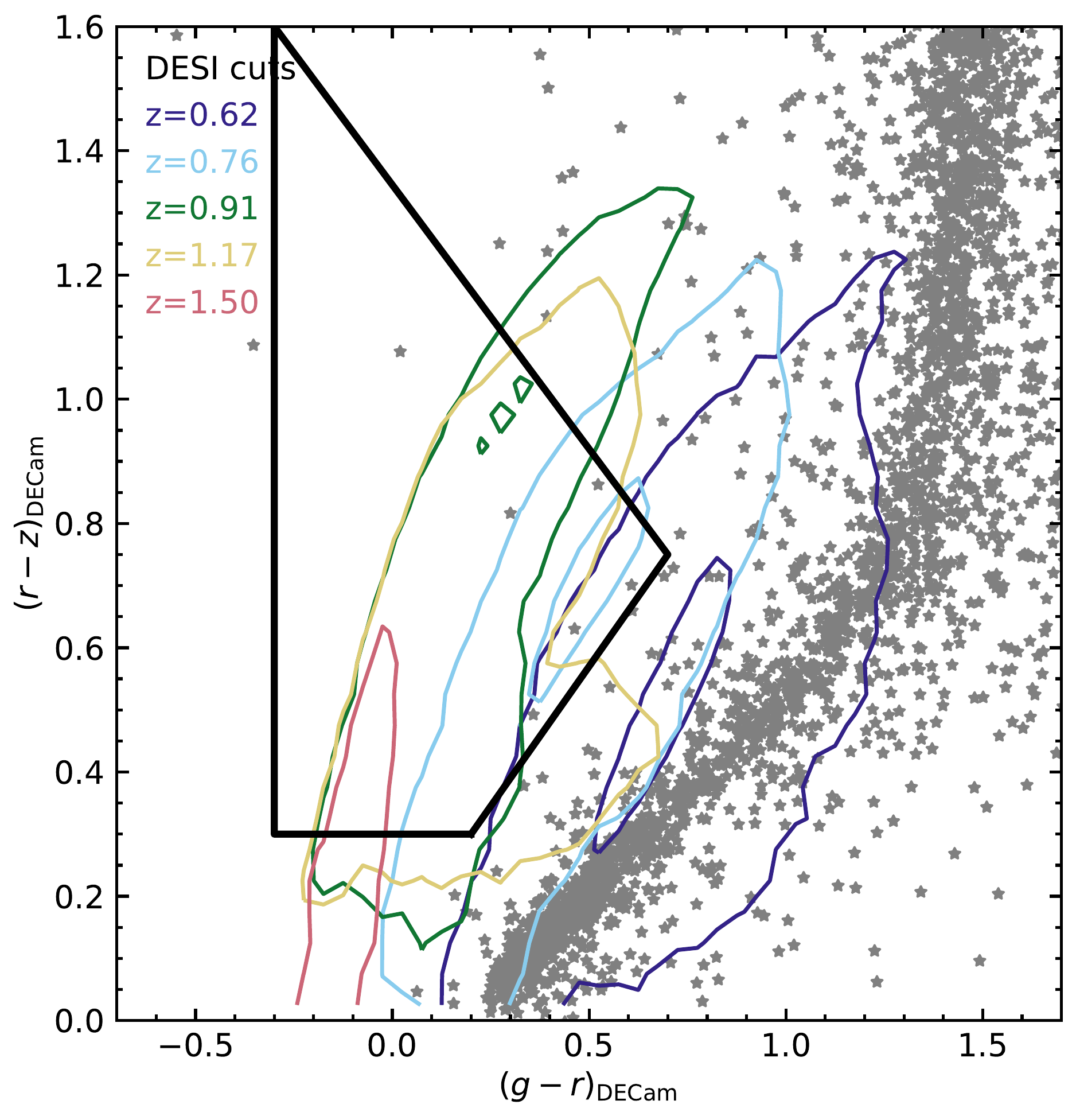}
\caption{\label{fig:colours} DECam (g-r) vs (r-z) parameter space with the isodensity lines at $log_{10}(\Phi/Mpc^{-3}h^{3}\rm{dlog_{10}}L)=-4.5, -1.5$ for model galaxies, with ${\rm Flux}_{\rm [OII]}>10^{-16}{\rm erg\, s}^{-1}{\rm cm}^{-2}$ and $22.1< g_{AB}^{\rm DECam} < 22.8$ (top panel) and ${\rm Flux}_{\rm [OII]}>8\cdot 10^{-17}{\rm erg\, s}^{-1}{\rm cm}^{-2}$ and $r_{AB}^{\rm DECam} < 23.4$ (bottom panel) at the redshifts indicated in the legend. The polygons indicate the region enclosed by the eBOSS (top panel) and DESI colour cuts (bottom panel) as summarized in Table \ref{tbl:obs}. The grey symbols in both panels show the location of stars.
} 
\end{figure}

Fig. \ref{fig:colours} presents the location of model galaxies at redshifts $z=0.62, 0.76, 0.91, 1.17, 1.5$ in the $(g-r)_{\rm DECam}$ vs. $(r-z)_{\rm DECam}$, colour-colour space, compared to the regions delimited by the colour cuts {\it decam180} described in \citet{comparat15eboss} and summarized in Table~\ref{tbl:obs} as eBOSS, top panel, and the DESI selection \citep{desi1}, bottom panel. The top panel shows model galaxies with ${\rm Flux}_{\rm [OII]}>10^{-16}{\rm erg\, s}^{-1}{\rm cm}^{-2}$ and $22.1< g_{AB}^{\rm DECam} < 22.8$. This magnitude cut mostly removes red galaxies from the colour-colour plot in Fig. \ref{fig:colours}. The bottom panel shows model galaxies with ${\rm Flux}_{\rm [OII]}>8\cdot 10^{-17}{\rm erg\, s}^{-1}{\rm cm}^{-2}$ and $r_{AB}^{\rm DECam} < 23.4$. The colours of model galaxies are roughly consistent with the regions defined tentatively for eBOSS \citet{comparat15eboss} and DESI \citep{desi1} to select ELGs at $z\sim 1$. A more detailed comparison shows the model $(g-r)_{\rm DECAM}$ to be just about 0.2 magnitudes redder than the observations presented in \citet{comparat15eboss}.

In the selection of model galaxies we use total magnitudes. The difference with respect to SDSS model magnitudes is expected to be less than 10\% for most model galaxies \citep{gonzalez09}. 

Fig. \ref{fig:colours} also shows the location of stars in the $(g-r)_{\rm DECam}$ vs. $(r-z)_{\rm DECam}$ space \citep[data from][cross-matched with the COSMOS DR3 legacy survey]{stars}. These overlap with the region occupied by galaxies at $z=0.62$ and for the bluest galaxies at $0.6<z<1.6$. Both the eBOSS and DESI selections reported here, are trading off selecting high redshift galaxies while minimising the stellar contamination.


\bsp	
\label{lastpage}
\end{document}